# Large-scale fading behavior for a cellular network with uniform spatial distribution

Mouhamed Abdulla[1*] and Yousef R. Shayan[2]

[1] Department of Electrical Engineering, University of Québec, Montréal, Canada
[2] Department of Electrical and Computer Engineering, Concordia University, Montréal, Canada

## ABSTRACT

Large-scale fading (LSF) between interacting nodes is a fundamental element in radio communications, responsible for weakening the propagation, and thus worsening the service quality. Given the importance of channel-losses in general, and the inevitability of random spatial geometry in real-life wireless networks, it was then natural to merge these two paradigms together in order to obtain an improved stochastical model for the LSF indicator. Therefore, in exact closed-form notation, we generically derived the LSF distribution between a prepositioned reference base-station and an arbitrary node for a multi-cellular random network model. In fact, we provided an explicit and definitive formulation that considered at once: the lattice profile, the users' random geometry, the effect of the far-field phenomenon, the path-loss behavior, and the stochastic impact of channel scatters. The veracity and accuracy of the theoretical analysis were also confirmed through Monte Carlo simulations.

***Correspondence**

Mouhamed Abdulla, Department of Electrical Engineering, University of Québec, Montréal, Canada.

E-mail: ma14@ieee.org

Website: www.DrMoe.org

## 1. INTRODUCTION

For wireless communications, large-scale fading (LSF) is indeed a basic consequence of the signal propagation between a base-station (BS) and a mobile node. In fact, because of its prerequisite for a host of network metrics including outage probability, the probability density function (PDF) for the path-loss (PL) or the received power level has been previously shown for a fixed predetermined separation between a node and a BS [1–5]. The aim in this paper is to reconsider this analytical problem for a multi-cellular network (MCN) architecture by generalizing the channel-loss distribution between any uniformly-based random positioned node and a preassigned BS reference. Evidently, this PDF can typically be obtained experimentally based on Monte Carlo (MC) simulations. However, there are two reasons why this approach is inconvenient: (i) random simulation is computationally expensive; and (ii) the obtained result is analytically intractable. These factors are further testaments for the necessity to obtain an explicit, generic, and rigorous theoretical derivation for the LSF density.

In recent years, some relevant work in the direction of random uniform spatial distribution model has gradually emerged; this effort is chronicled as follows. Initially, the contribution of [6] found the PL density for uniformly deployed nodes in a fixed circular cell. Then, an attempt to simplify this density result through curve fitting was shown in [7]. Next, we generalized in [8] the previous analysis in order to ensure spatial adaptability for various disk-based surface regions, along with multi-width rings and circular sectors. Furthermore, we derived in [9] the exact LSF distribution for an MCN between a random node and a reference BS located at the centroid of an hexagonal cell. Following the publication of our paper, using a slightly different approach, another paper appeared that also determined the PL density within a hexagonal cell and provided approximate options [10]. However, these outcomes did not specifically take into account a comprehensive and precise analysis that incorporated at once: the structure of the network configuration, users' nodal geometry, the effect of the far-field phenomenon, the PL predictive behavior, and the impact of channel shadowing due to in-field scatterers. Thus, while remaining generic and scalable for different network purposes, we aim here to accurately and explicitly solve this challenge by holistically formulating the propagation fundamentals of the LSF model for an adaptable random MCN pattern.

The rest of this paper is organized as follows. In Section 2, we will set the stage for network analysis by jointly assimilating the fundamental characteristics of spatial uniformity, lattice geometry, and radiation modeling. Then, in Section 3, the efficient and unbiased random network emulation geared for channel analysis will be developed. Respectively, in Sections 4 and 5, the LSF distribution analysis will be derived, and the exact closed-form stochastic result will be verified using MC simulations. Finally, Section 6 will close the paper. The symbols used in the paper are listed in Tables I and II.

## 2. CHARACTERISTICS OF THE NETWORK MODEL

Despite various conjectures for reconstructing a network based on inhomogeneous techniques, e.g. [11-13], the random uniform distribution assumption has been considered in analytical research, for example [14–19]. Essentially, if we consider $A_0 \in \mathbb{R}^+$ to be the surface area of a particular network lattice, and $n_0 \in \mathbb{N}^*$ to represent the scale of the architecture, then the uniform areal density will be given by $\rho_0 \triangleq n_0/A_0$. Generally, this simple spatial realization is feasible when no major information about the network site is available.

As shown in Figure 1, geometrical changes to the emulated network model can be applied in order to simplify the analysis. Indeed, it is clearly possible to dismember the hexagonal cell into smaller repetitive forms. In fact, the equilateral triangle is the most elementary portion of this cell model. Thus, considering this sub-pattern for internodal analysis will alleviate the derivation complexity of the LSF distribution because the formulation only depends on the reference to mobile separation, and is unaffected by the

**Table I.** Notations and symbols used in the paper – part 1.

| Symbol | Definition/Explanation |
|---|---|
| $\mathbf{1}_A(x)$ | Indicator function where unity is the case if $x \in A \subseteq \mathbb{R}$ |
| $\alpha, \beta$ | PL parameters for a particular link $(\text{dB})$ |
| $\delta_X(x)$ | Comparison probability density function of $f_X(x)$ used for the ARM algorithm |
| $\theta$ | Angular coordinate for polar notation $(\text{rad})$ |
| $\vec{\Lambda}$ | Array of generic attributes for the LSF distribution |
| $\mu$ | Cellular radius to the close-in distance ratio $(\text{RCR})$ |
| $\mu_{\sigma_{\min}}, \mu_{\sigma_{\max}}$ | Minimum and maximum RCR values for the AR estimator variance |
| $\mu_I$ | RCR value at the intersection point between Cartesian and radial AR functions |
| $\mu_{opt}$ | Optimum RCR value for random generation |
| $\pi_b(x)$ | Arbitrary bounding function of $f_X(x)$ used for the ARM algorithm |
| $\rho_0$ | Areal number density of a random network $(\text{no./unit of area})$ |
| $\sigma_\Psi$ | Standard deviation of shadowing $(\text{dB})$ |
| $\chi$ | Set of arguments for the infimum of $f_X(x)$ |
| $\hat{\psi}$ | Random instance of shadowing $(\text{dB})$ |
| $\Psi_{S-dB}$ | Shadowing element that emulates in-field scatterers |
| $A_0$ | Surface area of a network lattice $(\text{unit of area})$ |
| $A_{FF}$ | Deployment area with the effect of far-field for LSF analysis $(\text{unit of area})$ |
| $Binomial(x, n, p)$ | Binomial PMF for getting $x$ successes in $n$ trials, where each successful event has probability $p$ |
| $D_{FF}, D_{FF}^P$ | Support domain for the deployment surface in Cartesian and polar formats |
| $erf(x), erfc(x)$ | Error function, and complementary error function |
| $f_0(\tau)$ | Integrand of the LSF distribution |
| $f_\Psi(l)$ | Distribution function of shadowing |
| $f_{L_{PL}}(l, \vec{\Lambda})$ | Generic PDF of the LSF measure |
| $f_R(r)$ | Distribution function of the internodal distance |
| $f_R^{\max}$ | Maximum value of the radial PDF |
| $f_{R\theta}(r, \theta)$ | Joint polar PDF of the spatial random network |
| $f_W(w)$ | Density of the average decay |
| $f_X(x)$ | Marginal PDF for random network geometry along the $x$-axis |
| $f_X^{\max}$ | Maximum value of the marginal PDF along the $x$-axis |
| $(F_X)^{-1}(\hat{u})$ | ICDF used to generate random geometrical instances along the $x$-axis $(\text{unit of length})$ |
| $f_{XY}(x, y)$ | Spatial density function of a network cluster in Cartesian coordinate system |
| $f_{Y\mid X=\hat{x}}(y)$ | Conditional PDF of a random network along the $y$-axis |

PL, path-loss; ARM, acceptance rejection method; LSF, large-scale fading; AR, acceptance rate; PDF, probability density function; ICDF, inverse cumulative distribution function; PMF, probability mass function.

**Table II.** Notations and symbols used in the paper – part 2.

| Symbol | Definition/Explanation |
|---|---|
| $k, k_{\min}$ | Constants that enlarge $\delta_X(x)$ |
| $L$ | Predefined size of the cellular radius (unit of length) |
| $l$ | Random sample of LSF between a reference and an arbitrary terminal (dB) |
| $\tilde{l}_0, \tilde{l}_L$ | Measures w.h.p. the lower and higher extremities of LSF for an $L$-sized cell (dB) |
| $\hat{\bar{l}}, \hat{l}$ | Random instance for the average PL and LSF between a reference and an arbitrary node (dB) |
| $\Delta l_B$ | Width of each histogram bin for estimating the LSF density (dB) |
| $L_{PL}(r)_{dB}$ | LSF level |
| $\overline{L_{PL}(r)}_{dB}$ | Average PL decay |
| $m_{N_S}, \sigma_{N_S}$ | Mean and standard deviation of random variable $N_S$ |
| $m_{\tilde{p}_A}, \sigma_{\tilde{p}_A}$ | Mean and standard deviation of estimator $\tilde{p}_A$ |
| $\hat{n}$ | Random instance from a standard Gaussian PDF |
| $n_0$ | Amount of random nodes enclosed by a network lattice or cluster |
| $n_B$ | Quantity of histogram bars considered for density estimation |
| $n_{PL}$ | PL exponent |
| $n_S$ | Amount of i.i.d. randomly generated samples or nodes |
| $N_S$ | Random variable representing the number of accepted samples |
| $n_T$ | Total number of randomly generated instances |
| $\mathcal{N}(m, \sigma^2)$ | Gaussian PDF with mean $m \in \mathbb{R}$ and standard deviation $\sigma \in \mathbb{R}^+$ |
| $\mathbb{N}^*$ | Set of non-zero natural numbers |
| $O(\blacksquare)$ | Big-O notation for assessing the growth rate |
| $p_A = \Pr\{A \subset \Omega\}$ | Probability for accepting a randomly generated sample in space $\Omega$ |
| $\tilde{p}_A$ | MC estimator for the acceptance probability of samples |
| $pdf_j, cdf_j$ | Estimated PDF and CDF value measured numerically at the $j$th bin |
| $Q(x)$ | Q-function, which is a variation of the error function |
| $r$ | Random sample of the interpoint distance (unit of length) |
| $\hat{r}$ | Instance of the interspace between the reference and a node (unit of length) |
| $r_0$ | Close-in distance of an omni-directional reference antenna (unit of length) |
| $\mathbb{R}^+$ | Set of positive real numbers |
| $\hat{u}$ | Sample occurrence generated from a standard uniform PDF |
| $\mathcal{U}(a,b)$ | Continuous uniform PDF bounded by $[a,b] \in \mathbb{R}^2$ |
| $w_0, w_L$ | Average channel-loss at the close-in distance and the cell border (dB) |
| $w(r)$ | Random variable for the average PL (dB) |
| $\hat{x}, \hat{y}$ | Geometrical occurrence generated for the coordinate pair of a random node $(\text{unit of length})^2$ |

LSF, large-scale fading; w.h.p., with high probability; PL, path-loss; PDF, probability density function; i.i.d., independent and identically distributed; MC, Monte Carlo; CDF, cumulative distribution function.

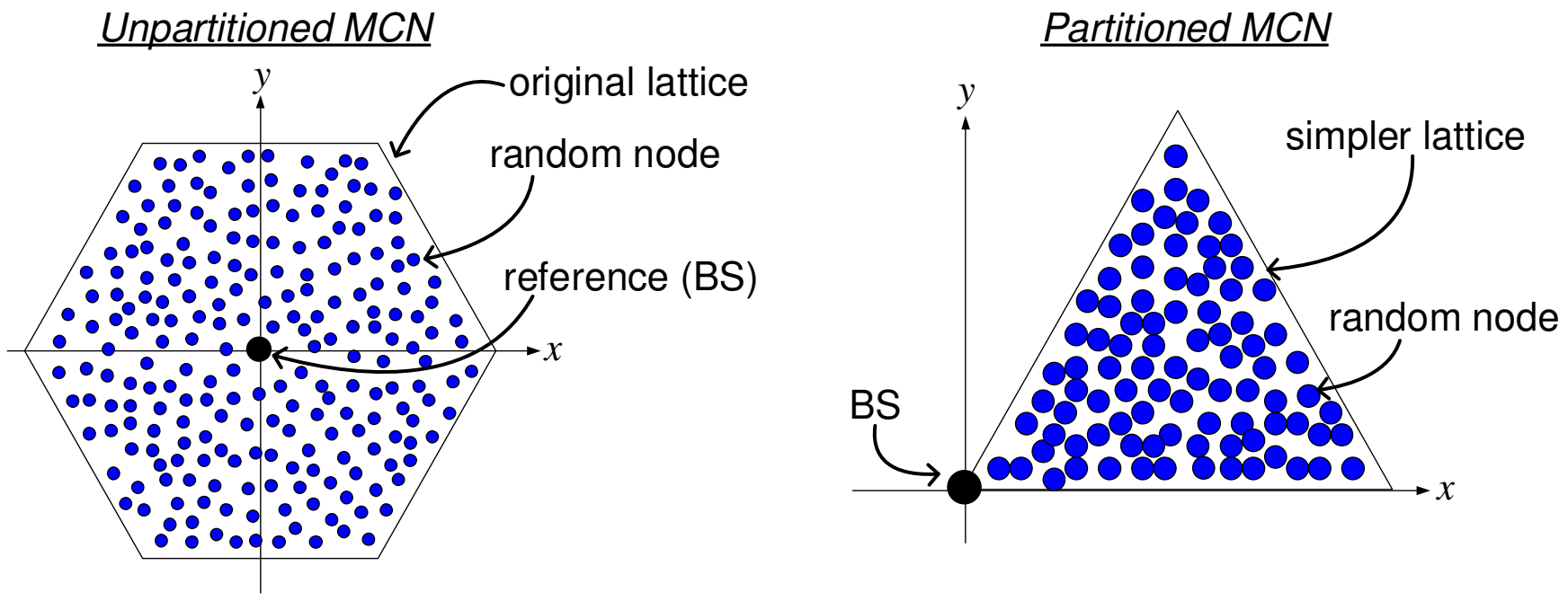

**Figure 1.** Simplifying channel analysis via geometrical partitioning. MCN, multi-cellular network; BS, base-station.

sectors rotation angle. Moreover, for planar deployment, the areal density will not affect the channel analysis because the network spatial distribution will remain *random* and *uniform*.

## 3. RANDOM NETWORK MODELING FOR ANALYSIS

### 3.1 Geometrical analysis

The characteristics described by nodal homogeneity, lattice geometry, and far-field radiation phenomenon, must collectively be incorporated in the spatial properties of the random network. In principle, this integration has a dual purpose: (i) it will be used to stochastically model the random lattice and effectively derive the PL density function for the entire network between a reference and an arbitrary terminal; and (ii) it will be employed to emulate actual random pattern instances, and numerically verify by means of MC simulations the precision of the anticipated LSF formulation.

To proceed, in Figure 2 the hexagonal cell is represented with the far-field region. In this surface model, the cellular size $L \in \mathbb{R}^+$ and the far-field limit $r_0 \in \mathbb{R}^+$ are the essential elements that define the entire geometry of the network structure. For notational convenience, we will define a parameter for the cellular radius to the close-in distance ratio (RCR): $\mu \triangleq L/r_0$ . From this model, we can determine the support range for the RCR indicator such that the layout of the lattice is accordingly preserved, namely, $\{r_0 < L/2\} \cap \{r_0 < \sqrt{3}L/2\} = \{\mu \in \mathbb{R}^+ | \mu > 2\}$ .

An expression in Cartesian coordinate notation for the spatial density function of a network cluster can be obtained via the deployment area, that is, $f_{XY}(x, y) = 1/A_{FF} = 12/\left(3\sqrt{3}L^2 - 2\pi r_0^2\right)$ . As for the marginal PDF for the nodal geometry along the *x*-axis, it can be computed as follows:

$$\begin{aligned} f_X(x) = \int_{(x,y)\in D_{FF}} f_{XY}(x,y)\,dy = \left\{12/\left(3\sqrt{3}L^2 - 2\pi r_0^2\right)\right\} \\ \times\left\{\left(\sqrt{3}x - \sqrt{r_0^2 - x^2}\right)\cdot \mathbf{1}\left(r_0/2 \le x \le r_0\right)\right. \\ + \sqrt{3}x \cdot \mathbf{1}\left(r_0 \le x \le L/2\right) \\ \left. + \sqrt{3}\left(L - x\right)\cdot \mathbf{1}\left(L/2 \le x \le L\right)\right\} \end{aligned} \quad (1)$$

### 3.2 Random spatial generation

The most efficient way to randomly generate arbitrary instances would be to consider the inverse transformation method (ITM), which is only possible through the use of the inverse cumulative distribution function (ICDF). In other words: $\hat{x} = \left\{(F_X)^{-1}\left(\hat{u} \sim \mathcal{U}(0,1)\right)\right\} \sim f_X(x)$ , where $\mathcal{U}(0,1)$ is a standard uniform distribution [20]. Clearly, the precondition in this approach requires the availability of the ICDF in explicit notation, which is actually impossible to achieve for the marginal density of (1). As an alternative, the acceptance rejection method (ARM) can be used for random number generation (RNG) [21]. Granted, this iterative process is suboptimal when compared to the ITM technique; nonetheless, we will develop an approach for modifying the ARM algorithm in order to maximize its performance.

Consider the distribution function $f_X(x): D_X \mapsto \mathbb{R}^+$, where the domain of the density is $D_X \triangleq [x_\alpha, x_\beta]$, and its associated extremities are given by $x_\alpha \triangleq \min \chi \in \mathbb{R}$ and $x_\beta \triangleq \max \chi \in \mathbb{R}$ such that $\chi \equiv \arg\{\inf(f_X(x) > 0)\} \subset (x \in \mathbb{R})$ . Then, based on the ARM procedure, we would need to determine some continuous arbitrary bounding function, say $\pi_b(x): D_X \mapsto \mathbb{R}^+$ , that covers the domain of $f_X(x)$, while $\pi_b(x) \ge f_X(x)$ . Moreover, this bounding function is expected to be an augmented version to some valid comparison PDF $\delta_X(x): D_X \mapsto \mathbb{R}^+$ . In fact, the most generic and simplest way would be to consider the uniform case for the comparison density, namely, $\delta_X(x) = \mathcal{U}_X(x_\alpha, x_\beta)$. And thus, the bounding function can be realized by $\pi_b(x) = k\delta_X(x) \quad \exists k \in \mathbb{R}^+ : k \geq 1$. Meanwhile, the likelihood for accepting a randomly generated sample is specified by the area below $f_X(x)$. In contrast, the remaining sector between $\pi_b(x)$ and $f_X(x)$ constitutes the rejection region of generated samples. In order to maximize the acceptance rate (AR) of arbitrary samples, we could in essence minimize the rejection region. This could for instance be leveraged by adjusting the growth constant $k$ to $k_{\min}$, such that $k > k_{\min} \geq 1$. To obtain this element, we need to identify the maximum value of the PDF: $f_X^{\max} \triangleq \max_{x \in \mathbb{R}} \{f_X(x)\} \in \mathbb{R}^+$. Then, we perform the following association: $\inf_{k \in \mathbb{R}^+} \{\pi_b(x)\} = f_X^{\max}$, and so we realize that: $k_{\min} = f_X^{\max} / \delta_X(x) = f_X^{\max} / \mathcal{U}_X(x_\alpha, x_\beta) = f_X^{\max}(x_\beta - x_\alpha)$. Next, a decision for the suitability of a sample for random generation based on the ARM algorithm depends on the $f_X(\hat{v})/\pi_b(\hat{v})$ ratio, where $\hat{v} \sim \delta_X(x)$. This expression can further be elaborated as follows:

$$\frac{f_X(\hat{v})}{\pi_b(\hat{v})} = \frac{f_X(\hat{v})}{k\delta_X(\hat{v})} < \frac{f_X(\hat{v})}{k_{\min}\delta_X(\hat{v})} = \frac{f_X(\hat{v})}{f_X^{\max}(x_\beta - x_\alpha)\mathcal{U}_X(x_\alpha, x_\beta)} = \frac{f_X(\hat{v})}{f_X^{\max}} \tag{2}$$

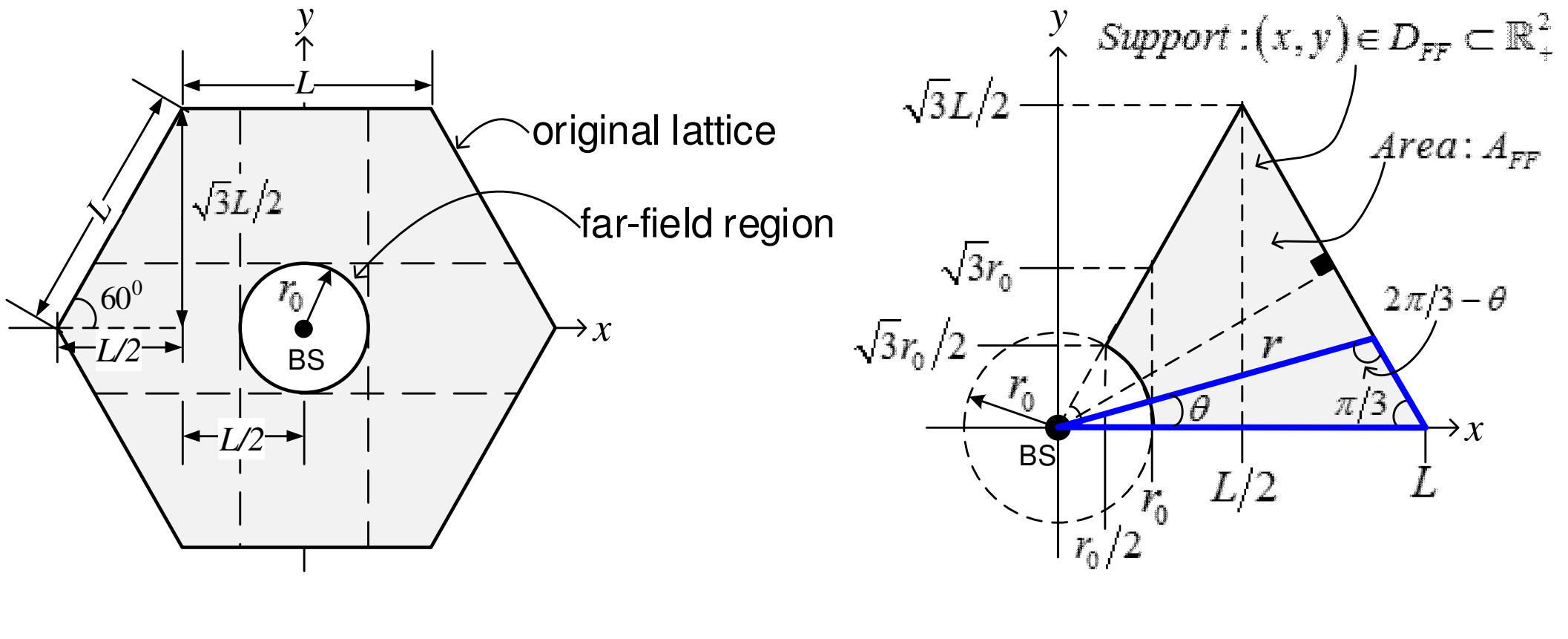

**Figure 2.** Dimensions of the random network deployment surface with far-field. BS, base-station.

If we apply (2) to the marginal PDF of (1), we then obtain:

$$\begin{aligned} \frac{f_X(\hat{v})}{f_X^{\max}} = 2\Big\{ & \left(\hat{v} - \sqrt{\left(r_0^2 - \hat{v}^2\right)/3}\right) \Big/ L \cdot \mathbf{1}\left(r_0/2 \leq \hat{v} \leq r_0\right) \\ & + \hat{v}/L \cdot \mathbf{1}\left(r_0 \leq \hat{v} \leq L/2\right) \\ & + (1 - \hat{v}/L).\mathbf{1}\left(L/2 \leq \hat{v} \leq L\right)\Big\} \end{aligned} \tag{3}$$

After taking the above analysis into account, we then obtain the RNG algorithm for $\hat{x} \sim f_X(x)$ in Figure 3 that ensures an efficient approach for generating $n_S$ samples.

**Algorithm 1** - Random Deployment with Far-Field Radiation along the $x$-axis

```
Require: n_S ∈ ℕ*   r_0 ∈ ℝ+   L ∈ ℝ+
Initialize: i = 0   n_T = 0
while {i < n_S} do
    n_T := n_T + 1
    Generate two i.i.d. RVs: {û_0, û_1} ~ 𝒰(0,1)
    Compute: v̂ = r_0/2 + û_1(L − r_0/2) ~ δ_X(x) = 𝒰_X(r_0/2, L)
    if {r_0/2 ≤ v̂ ≤ r_0} then
        if {v̂ − sqrt((r_0^2 − v̂^2)/3) > û_0 L/2} then
            i := i + 1
            Random sample is accepted: x̂_i := v̂ ~ f_X(x)
        end if
    else if {r_0 ≤ v̂ ≤ L/2} then
        if {v̂ > û_0 L/2} then
            i := i + 1
            Random sample is accepted: x̂_i := v̂ ~ f_X(x)
        end if
    else if {L/2 ≤ v̂ ≤ L} then
        if {v̂ < L(1 − û_0/2)} then
            i := i + 1
            Random sample is accepted: x̂_i := v̂ ~ f_X(x)
        end if
    end if
end while
Return/Compute: {x̂_i} : i = 1, 2, ···, n_S   p̃_A = n_S/n_T
```

**Figure 3.** Pseudocode for efficient random generation. RV, random variable.

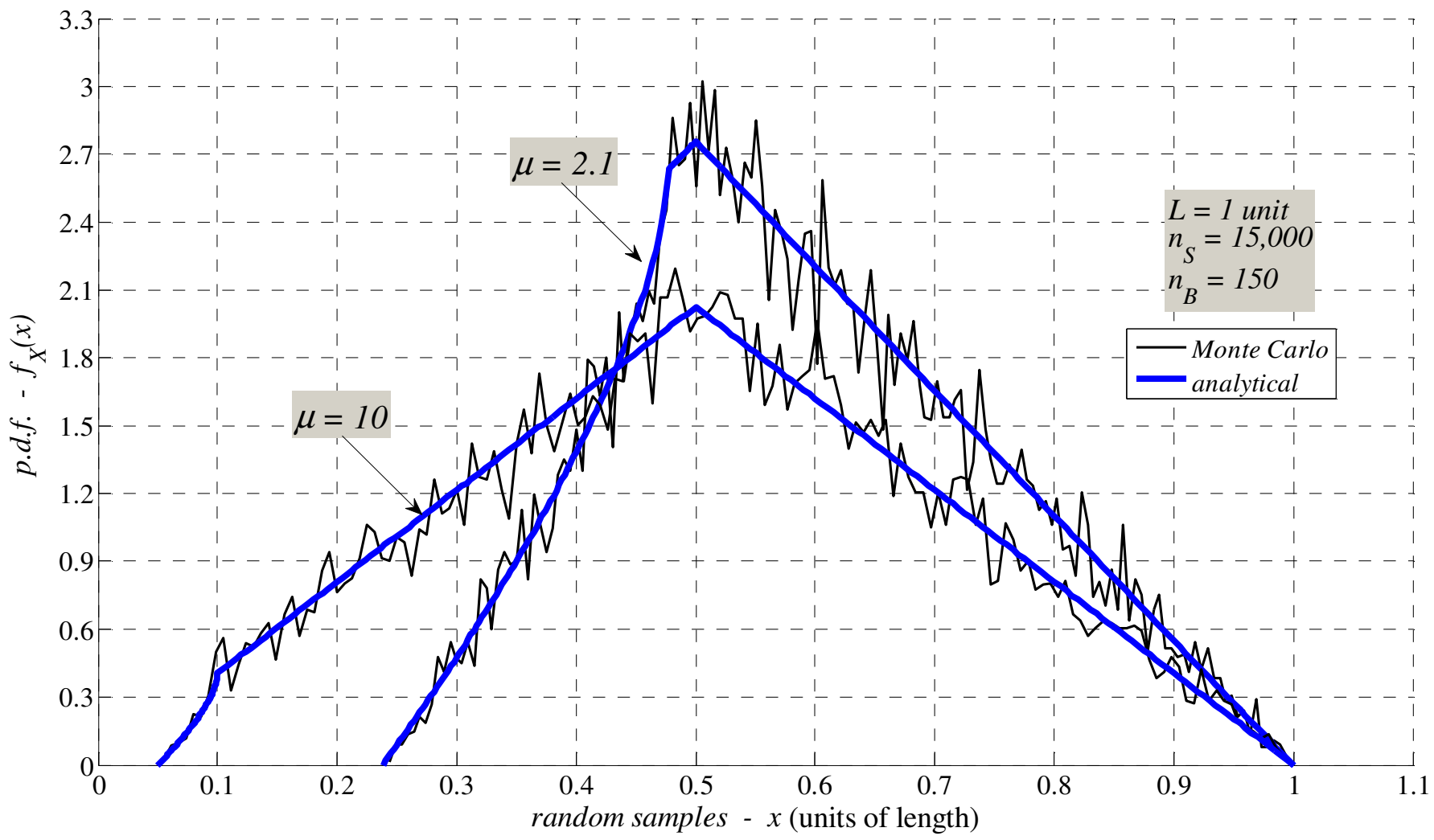

**Figure 4.** Marginal density of nodal geometry by means of random simulations. PDF, probability density function.

In Figure 4, the PDF along the $x$-axis is shown for two different values of RCR obtained by means of analysis and via MC simulations for $n_S = 15,000$ valid samples and with a histogram of $n_B = 150$ bins.

### 3.3 Measuring the performance of efficient random generation

In this part, we are interested to quantify the performance of the obtained efficient RNG. Thus, we want to determine an expression for the corresponding AR. Following the logic detailed previously, the event of accepting a sample is defined as a subset of the universal space $\Omega = \{A, R\}$. Consequently, the AR as a function of RCR can be determined by (4).

$$p_A = \Pr\{A \subset \Omega\} = \int_{x=-\infty}^{\infty} f_X(x)\,dx \Big/ \int_{x=-\infty}^{\infty} \pi_b(x)\,dx = 1\Big/\int_{x=-\infty}^{\infty} k\delta_X(x)\,dx = 1\Big/ k_{\min}\int_{x=-\infty}^{\infty} \mathcal{U}_X(x_\alpha, x_\beta)\,dx$$
$$= 1/k_{\min} = 1\Big/ f_X^{\max}(x_\beta - x_\alpha) = \left(L^2 - 2\pi r_0^2/3\sqrt{3}\right)\Big/ L(2L - r_0) = \left(\mu^2 - 2\pi/3\sqrt{3}\right)\Big/\mu(2\mu - 1) \quad \mu > 2 \tag{4}$$

In (4), we theoretically derived an expression for the AR. Conversely, we may also define a MC estimator for the acceptance probability of samples numerically assessed by $\tilde{p}_A = n_S/n_T$ such that $n_S \in \mathbb{N}$ represents the number of accepted samples and $n_T \in \mathbb{N}^*$ is the total number of randomly generated instances for a particular simulation realization. Assuming that $n_T$ is deterministic, then the number of accepted samples will be random with distribution, $N_S \sim Binomial(n_S, n_T, p_A)$, having mean and variance equal, respectively, to $m_{N_S} = n_T p_A$ and $\sigma_{N_S}^2 = n_T p_A(1 - p_A)$. Then the statistics of the AR estimator can be shown to equal:

$$m_{\tilde{p}_A} = m_{\tilde{p}_A}(\mu) = \mathrm{E}[\tilde{p}_A] = \mathrm{E}[N_S/n_T] = m_{N_S}/n_T = p_A$$
$$= \left(\mu^2 - 2\pi/3\sqrt{3}\right)\Big/\mu(2\mu - 1) \tag{5}$$

At this point, the natural intrigue is to analytically obtain the optimum RCR value that maximizes $p_A = p_A(\mu)$, which can be obtained by $dp_A(\mu)/d\mu = 0$; therefore resulting in a unique feasible solution given by: $\mu_{opt} = \left\{4\pi + \sqrt{2\pi\left(8\pi - 3\sqrt{3}\right)}\right\}\Big/3\sqrt{3} \approx 4.57$. Hence, the efficient random generation approach developed can further be improved when $\mu = \mu_{opt}$, which essentially ensures an AR of $p_A(\mu_{opt}) \approx 0.529$. Pursuing this further, it is also worthwhile to characterize the AR as the RCR progressively increases, which can be evaluated by: $\lim_{\mu\to\infty} p_A(\mu) = 1/2$. In fact, it can be shown that $p_A = 0.5$ is indeed a horizontal asymptote (HA) of the $p_A(\mu)$ function.

$$\sigma_{\tilde{p}_A}^2 = \sigma_{\tilde{p}_A}^2(\mu, n_T) = \mathrm{E}\left[\left(\tilde{p}_A - m_{\tilde{p}_A}\right)^2\right] = \mathrm{E}\left[\left(N_S/n_T - m_{\tilde{p}_A}\right)^2\right]$$
$$= \sigma_{N_S}^2/n_T^2 = p_A(1 - p_A)/n_T$$
$$= \left\{1 - \left(\mu - 4\pi/3\sqrt{3}\right)^2\Big/\mu^2(2\mu - 1)^2\right\}\Big/4n_T \tag{6}$$

Therefore, from (5) we realize that the AR estimator is *unbiased*, and from (6) we note that it is *consistent* because $\lim_{n_T\to\infty} \sigma_{\tilde{p}_A}^2 = 0$; meaning that an increase in $n_T$ will improve this estimator at the expense of running time complexity. Furthermore, it is desired to minimize the variance of the AR estimator in order to enhance its predictability.

Indeed, through the optimization of $\partial\sigma^2_{\tilde{p}_A}(\mu,n_T)/\partial\mu = 0$ and the plot of Figure 5, we determine that $\mu_{\sigma_{\min}} = \mu_{opt}$ is a feasible stationary point, and also detect a HA at $\sigma^2_{\tilde{p}_A}(\infty,n_T) = 1/4n_T$. Overall, we remark that selecting $\mu \approx 4.57$ has a dual statistical advantage: (i) it maximizes the AR for RNG; and (ii) it minimizes the AR estimator variance.

For further insight of the AR behavior, in Figure 6, we show the corresponding theoretical and experimental plots. In fact, during MC simulations, for each $\mu$ value, $n_S = 10,000$ accepted samples are sought to estimate the AR. Overall, this RNG approach is reasonably similar to a coin toss for all possible realization because the AR of the efficient algorithm in Figure 3 is confined to: $0.47 < p_A(\mu) < 0.53 \quad \mu \in (2,\infty)$.

### 3.4 Geometrical deployment on the Euclidian plane

Efficient deployment along the *x*-axis was developed in the previous subsections. In this part, we will extend the treatment by deriving the spatial emplacement along the *y*-axis in order to generate random coordinates on the Euclidian plane. To do this, we require the conditional PDF which we obtain by the use of (1) alongside the deployment support of Figure 2, which produces:

$$\begin{aligned} f_{Y|X=\hat{x}}(y) &= f_{XY}(\hat{x},y)/f_X(\hat{x}) \\ &= \mathcal{U}_Y\left(\sqrt{r_0^2-\hat{x}^2},\sqrt{3}\hat{x}\right)\cdot\mathbf{1}(r_0/2\le\hat{x}\le r_0) \\ &\quad+\mathcal{U}_Y\left(0,\sqrt{3}\hat{x}\right)\cdot\mathbf{1}(r_0\le\hat{x}\le L/2) \\ &\quad+\mathcal{U}_Y\left(0,\sqrt{3}(L-\hat{x})\right)\cdot\mathbf{1}(L/2\le\hat{x}\le L) \end{aligned} \tag{7}$$

In essence, depending on a particular sampling range for $\hat{x}$, the related PDF is then considered in the expression of (7) for randomly emulating the *y*-component of an arbitrary node. On the whole, the deployment complexity for the optimum spatial random generation can be assessed by integrating the algorithm of Figure 3 and the result formulated in (7) together: $O(n_T)+O(n_S)\sim O(n_S)$. Namely, the deployment of $n_S$ random terminals has a computational cost of $O(n_S)$ provided $\mu \gg 2$. At last, to geometrically demonstrate the analysis reported in this section, we simulated in Figure 7 the random deployment for different nodal scales and RCR values.

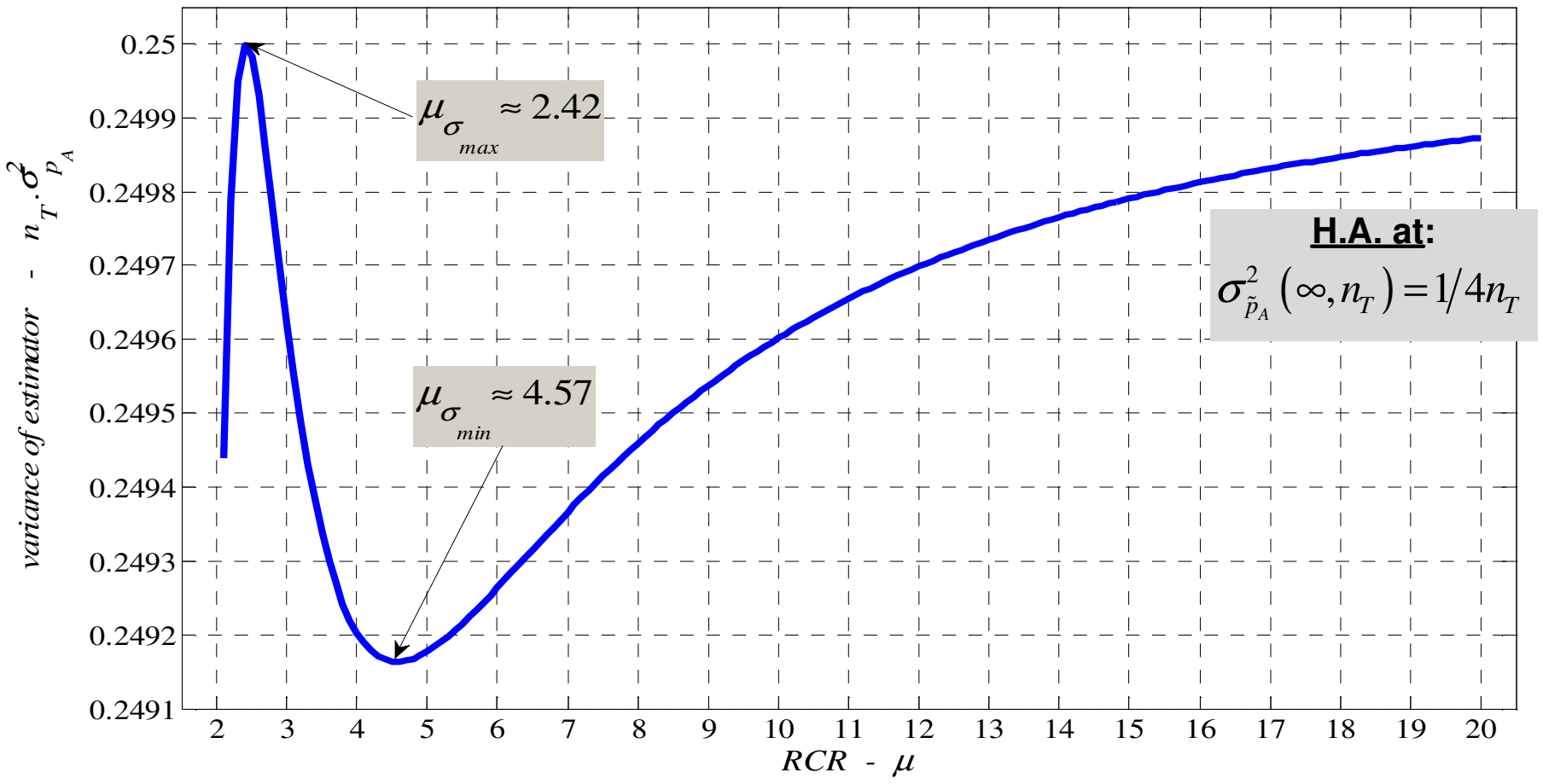

**Figure 5.** Impact of radius to the close-in distance ratio (RCR) on the acceptance rate estimator variance.

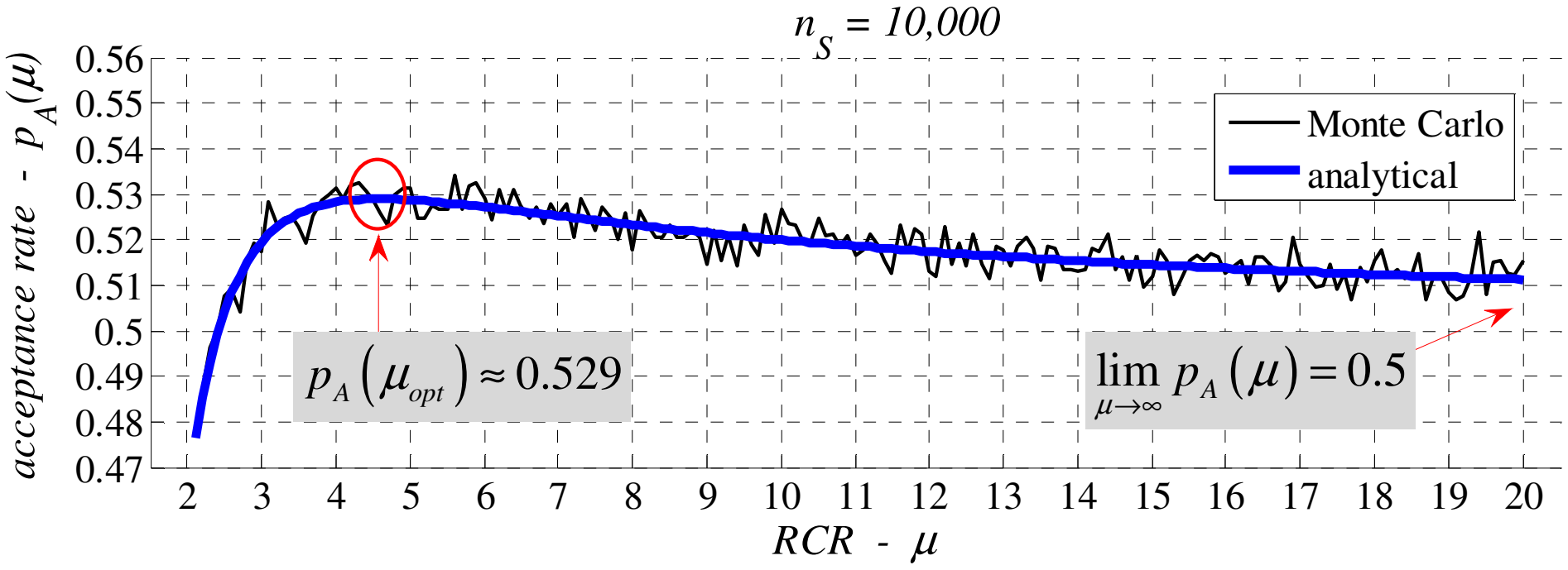

**Figure 6.** Acceptance rate for efficient random generation versus radius to the close-in distance ratio (RCR).

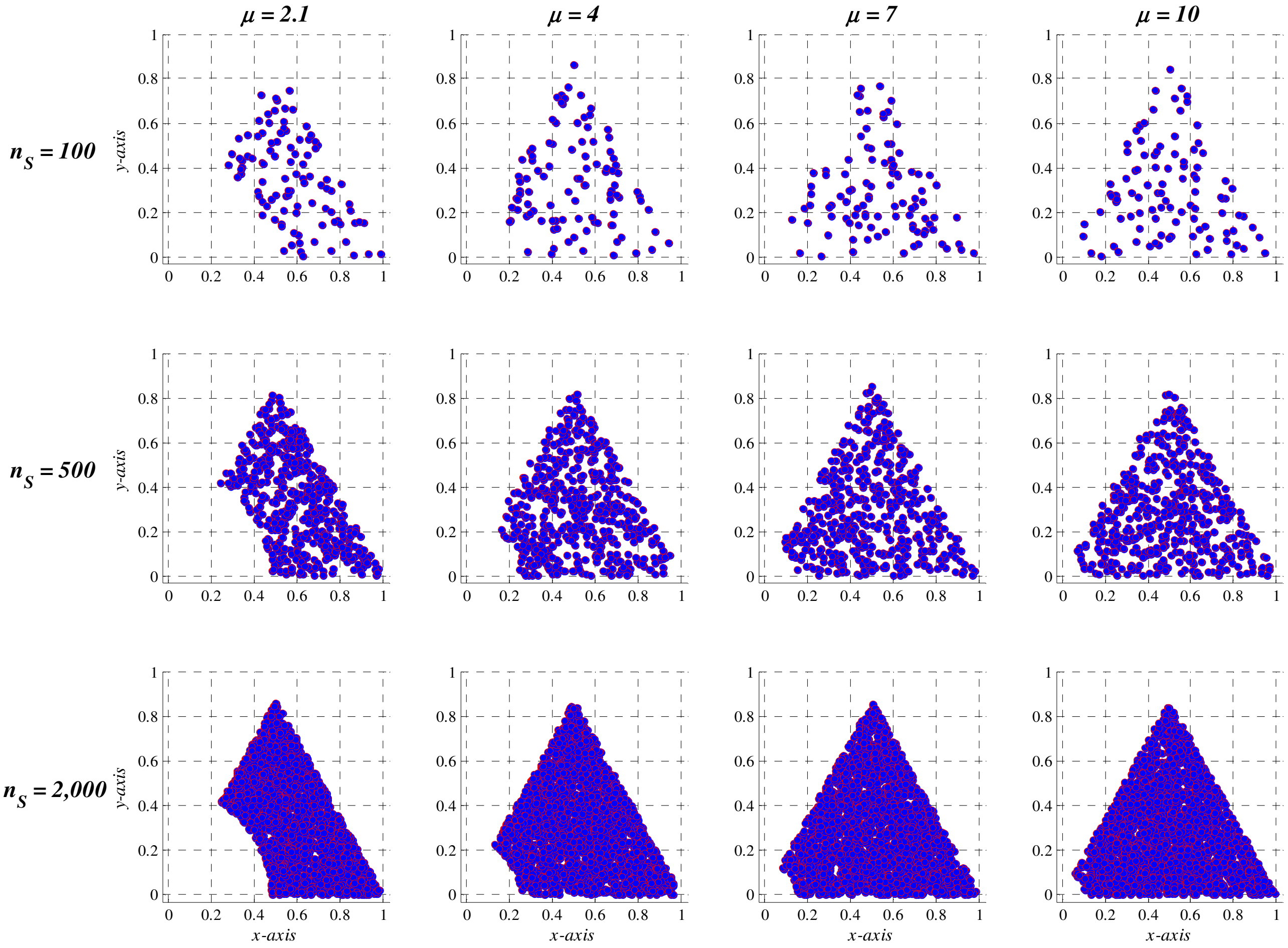

**Figure 7.** Random spatial emulation as a function of network scale and radius to the close-in distance ratio values.

## 4. LARGE-SCALE FADING ANALYSIS

### 4.1 Spatial density model in polar notation

The spatial behavior elaborated in the previous section was performed as groundwork for general network emulation, formulation of the LSF density, and to numerically verify the authenticity of the analysis. In this part, we are interested to move forward by describing the stochastic characteristics of the channel-loss between an arbitrary node and a reference located at the origin of the service area. Given the nature of this problem, analysis in polar notation is favored, thus the joint density changes to:

$$f_{R\theta}(r,\theta) = f_{XY}(x,y)\Big|_{\substack{x=r\cos\theta \\ y=r\sin\theta}} \cdot \left|\det\begin{pmatrix} \partial x/\partial r & \partial x/\partial\theta \\ \partial y/\partial r & \partial y/\partial\theta \end{pmatrix}\right|$$
$$= 12\cdot r\Big/\left(3\sqrt{3}L^2 - 2\pi r_0^2\right) \quad (r,\theta)\in D_{FF}^{P} \subset \mathbb{R}_+^2 \tag{8}$$

Using the law of sines to the marked blue triangle shown in Figure 2, an expression for the coverage radius can be obtained by $r_0 \le r \le r(\theta) = \sqrt{3}L\big/2\sin\left(2\pi/3-\theta\right)$ over $0 \le \theta \le \pi/3$. And therefore, the associated polar-based domain $D_{FF}^{P}$ can be formulated as follows:

$$D_{FF}^{P} = \left\{ \begin{array}{l|l} (r,\theta)\in\mathbb{R}_+^2; & 0\le\theta\le\pi/3 : r\in\left[r_0, \sqrt{3}L/2\right]; \\ (r_0, L)\in\mathbb{R}_+^2 : & 0\le\theta\le\arcsin\left(\sqrt{3}L/2r\right)-\pi/3 : r\in\left[\sqrt{3}L/2, L\right]; \\ r_0 < L/2 & 2\pi/3-\arcsin\left(\sqrt{3}L/2r\right)\le\theta\le\pi/3 : r\in\left[\sqrt{3}L/2, L\right] \end{array} \right\} \tag{9}$$

### 4.2 Characterizing radial distribution

In past contributions, the interpoint PDF has been shown between a fixed reference at the vertex of a triangle and a random point [22], the centroid of a polygon and a random point [23], and more generally among two arbitrary nodes inside a polygon [24]. Meanwhile, for the punctured hexagonal region of Figure 2, using (8) and (9), the radial PDF between a BS and a node can be obtained by:

$$f_R(r) = \int_{(r,\theta)\in D_{FF}^P} f_{R\theta}(r,\theta)\,d\theta = \left\{4\pi r \big/ \left(3\sqrt{3}L^2 - 2\pi r_0^2\right)\right\}\cdot \mathbf{1}\left(r_0 \le r \le \sqrt{3}L/2\right) + \left\{8r\left\{3\arcsin\left(\sqrt{3}L/2r\right) - \pi\right\} \big/ \left(3\sqrt{3}L^2 - 2\pi r_0^2\right)\right\}\cdot \mathbf{1}\left(\sqrt{3}L/2 \le r \le L\right) \quad (10)$$

This interpoint PDF can then be substantiated via the simulation results shown in Figure 8, where the theoretical and MC plots for a unity cell are accordingly graphed over two RCR values. In principle, for a particular $\mu$ value, the spatial position of $n_S = 25{,}000$ random nodes is generated in a manner similar to that carried in Figure 7. Then, the measure from the arbitrary node to the BS is computed and an $n_B = 250$ bin histogram is constructed and accordingly scaled for plotting the PDF.

### 4.3 RNG based on radial distribution

Having $f_R(r)$ leads us to appropriately remark that in order to verify the anticipated analytical formulation for LSF density, random MC data can also be generated straight from the radial distribution in addition to the Cartesian-based RNG analysis described in Section 3. To contrast the computational suitability of this generation option, we thus need to identify the RNG attributes of the radial PDF. It can in fact be shown that the most efficient ITM approach is unsuitable given that a closed-form ICDF is unattainable. As a workaround, the modified version of the ARM procedure can be considered for enhancing the generation performance of the radial probability distribution. Following the notation derived in (4), the utmost AR for the modified iterative algorithm becomes:

$$p_A^{Radial}(\mu) = 1\big/ f_R^{\max}\left(r_\beta - r_\alpha\right) = 1\big/ \max_{r\in\mathbb{R}}\left\{f_R(r)\right\}\cdot\left(L - r_0\right) = 3\left(\mu^2 - 2\pi/3\sqrt{3}\right)\big/ 2\pi\mu(\mu-1) \qquad \mu > 2 \quad (11)$$

Additionally, we can find the intersection point for the AR among the Cartesian and radial notations, which is located at: $\mu_I = (2\pi-3)/2(\pi-3) \approx 11.59$. For comparison purposes, in Figure 9, we graph the AR for both of these RNG approaches. As shown, the AR for the radial distribution is monotonically decreasing, whereas the Cartesian alternative is not monotonic at all. Moreover, the HA of (11), which equals to $\lim_{\mu\to\infty} p_A^{Radial}(\mu) = 3/2\pi \approx 0.48$, reveals that Cartesian-based RNG is more performant as the RCR extends beyond $\mu_I$. Overall, the optimum generation approach can thus be improved by partitioning the RCR range such that the AR is maximized. This leads us to observe the following association for further improvement to efficient random generation:

$$\begin{aligned} 2 < \mu \le \mu_I &\quad \leftrightarrow \quad \textit{radial RNG} \\ \mu > \mu_I &\quad \leftrightarrow \quad \textit{Cartesian RNG} \end{aligned} \quad (12)$$

### 4.4 Distribution of the average path-loss

In general, it is shown (say [1]) that the average PL for mobile cellular communications is modeled by $\overline{L_{PL}(r)} = \overline{L_{PL}(r_0)} \cdot (r/r_0)^{n_{PL}}$, where $n_{PL} \in \mathbb{R}^+ : n_{PL} > 1$ is the PL exponent, and $r_0, r \in \mathbb{R}_+^2 : r_0 \le r$ denote, respectively, the close-in distance and the internodal gap. For analytical suitability, this expression may be mapped to simpler notations, where the average PL for an $L$ sized cellular network at a generic internodal gap is characterized by $w(r) \equiv \overline{L_{PL}(r)}_{dB} = \alpha + \beta \log_{10}(r)$ over $0 < r_0 \le r \le L$. Also, its inverse, which will be required in the next step, equals to $r(w) = 10^{(w-\alpha)/\beta}$ where $w \in [w_0, w_L]$ are breakpoints interrelated to (10).

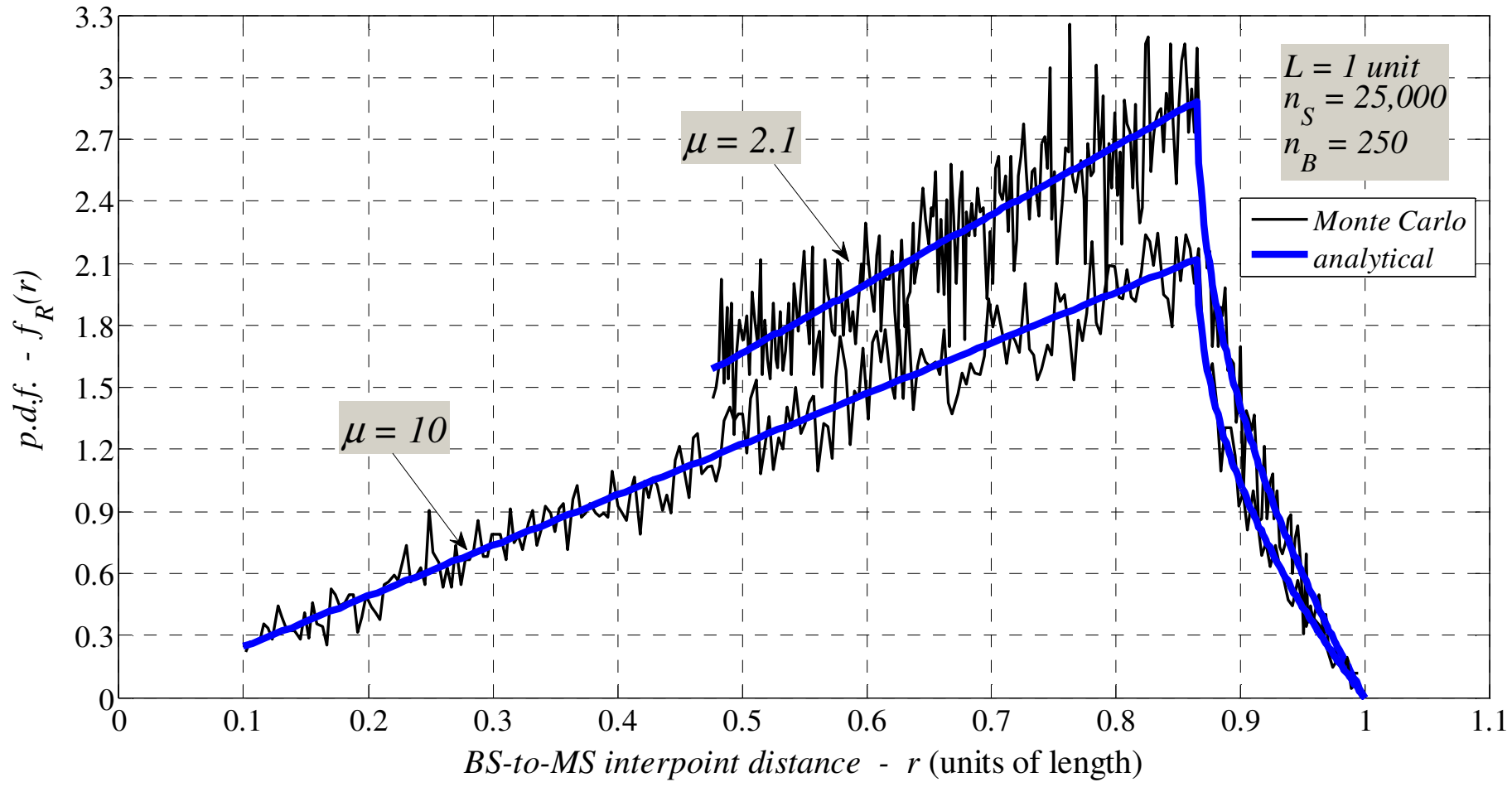

**Figure 8.** Radial distribution for nodal geometry via stochastic simulations. PDF, probability density function; BS, base-station; MS, mobile-station.

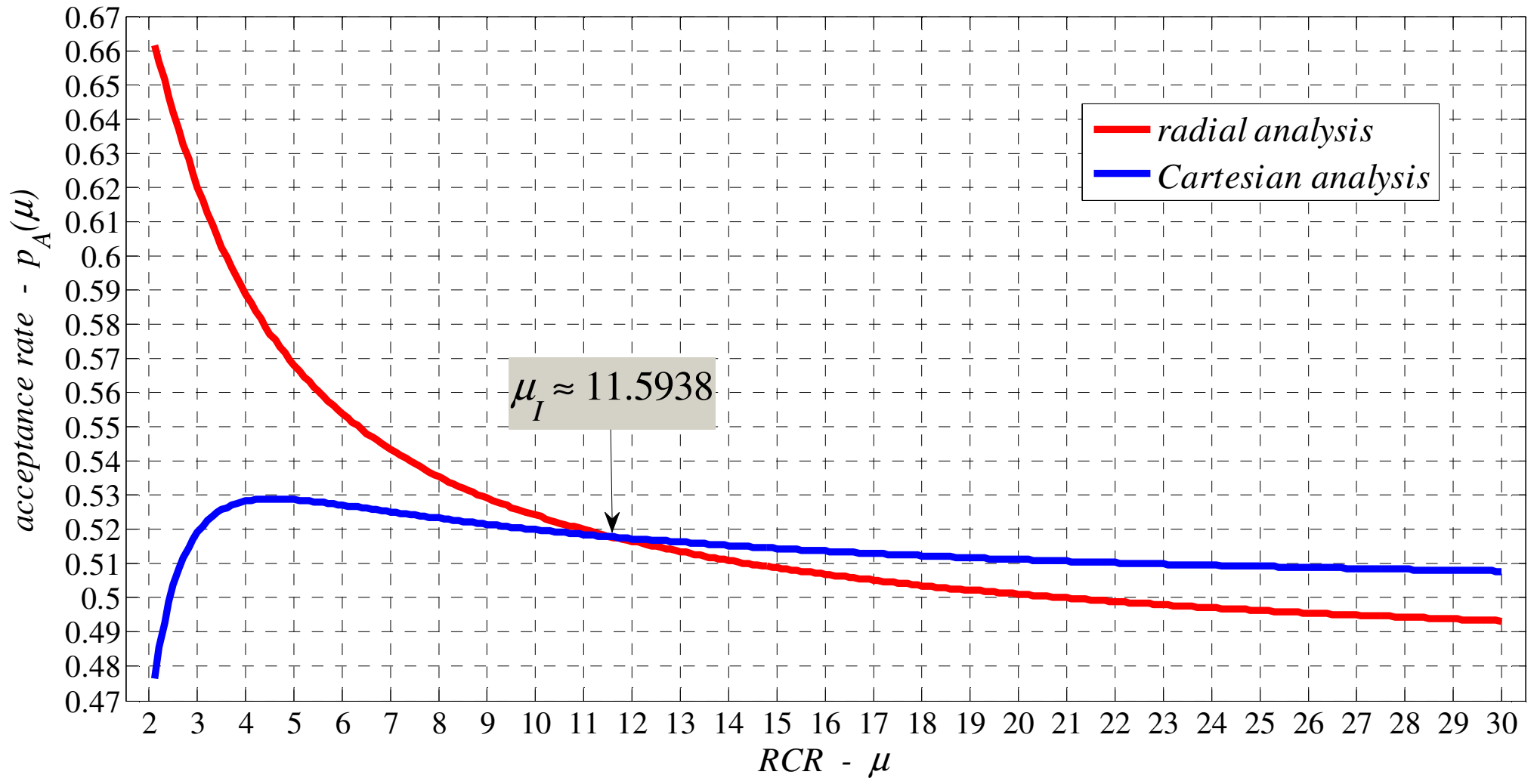

**Figure 9.** Efficient acceptance rate for random number generation based on radial and Cartesian analysis. RCR, radius to the close-in distance ratio.

The objective now is to characterize the distribution of the average PL overlaying the randomness of nodal geometry; therefore, we perform the following stochastic transformation:

$$w = w(r \sim f_R(r)) \equiv \overline{L_{PL}(r)}_{dB} \sim f_W(w) = f_R(r = r(w)) \Big/ \left| dw(r)/dr \right|_{r=r(w)}$$

$$\therefore f_W(w) = \frac{f_R\left(10^{(w-\alpha)/\beta}\right)}{\left|\dfrac{\beta}{\ln(10)\, r}\right|_{r=10^{(w-\alpha)/\beta}}} = \frac{\ln(10)\cdot 10^{(w-\alpha)/\beta}}{\beta} \left\{ \begin{aligned} &\frac{4\pi r}{\left(3\sqrt{3}L^2 - 2\pi r_0^2\right)} \cdot \mathbf{1}\left(r_0 \le r \le \sqrt{3}L/2\right) \\ &+ \frac{8r\left\{3\arcsin\left(\sqrt{3}L/2r\right) - \pi\right\}}{\left(3\sqrt{3}L^2 - 2\pi r_0^2\right)} \cdot \mathbf{1}\left(\sqrt{3}L/2 \le r \le L\right) \end{aligned} \right\}_{r=10^{(w-\alpha)/\beta}}$$

$$= \frac{4 \cdot \ln(10) \cdot 10^{2(w-\alpha)/\beta}}{\beta\left(3\sqrt{3}L^2 - 2\pi r_0^2\right)} \left\{ \begin{aligned} &\pi \cdot \mathbf{1}(w_0 \le w \le w_I) - 2\pi \cdot \mathbf{1}(w_I \le w \le w_L) \\ &\quad + 6\arcsin\left(\sqrt{3}L/2 \cdot 10^{(w-\alpha)/\beta}\right) \cdot \mathbf{1}(w_I \le w \le w_L) \end{aligned} \right\} \tag{13}$$

Basically, shadowing is accounted for by merely adding a random variable (RV) $\Psi_{S-dB}$ to the average PL. It is imperative to note that the randomness of shadowing and the average PL are statistically uncorrelated. Therefore, the overall LSF distribution is obtained by convolving the corresponding density functions:

$$\begin{aligned} L_{PL}(r)_{dB} &= \left\{\overline{L_{PL}(r)}_{dB} + \Psi_{S-dB}\right\} \cdot \mathbf{1}(r_0 \le r \le L) \sim f_{L_{PL}}(l) \\ &= (f_W * f_\Psi)(l) \\ \therefore f_{L_{PL}}(l) &= \int_{\tau=-\infty}^{\infty} \underbrace{f_W(\tau)}_{\text{path-loss}} \cdot \underbrace{f_\Psi(l-\tau)}_{\text{shadowing}} d\tau \triangleq \int_{\tau=-\infty}^{\infty} f_0(\tau)\, d\tau \\ & l \in \mathbb{R} \end{aligned} \tag{14}$$

## 4.5 Large-scale fading density with shadowing

In this part, we will supplement the PDF for the average power loss by introducing the impact of shadowing. In fact, this critical component analytically characterizes the implication of scatterers in the propagation channel; thus, incorporating it in the PL model is of paramount importance.

The shadowing entity is actually described by a zero-mean log-normal distribution with standard deviation (SD) $\sigma_\Psi$; i.e.: $\Psi_{S-dB} \sim f_\Psi(\tau) = \mathcal{N}_S(0, \sigma_\Psi^2)$. And therefore, with some analysis, it can be demonstrated that $f_\Psi(l-\tau) = \mathcal{N}_S(l, \sigma_\Psi^2) : \tau \in \mathbb{R},\ l \in \mathbb{R}^+$. As for the $f_W(\tau)$ part in (14), it is obtained by the notation in (13) following an exchange of $w$ by $\tau$. Consequently, the integrand of (14) reduces to the expression in (15), and having a domain which is limited by: $\{\tau \in \mathbb{R}\} \cap \{0 < w_0 \le \tau \le w_L\} = \{\tau \in \mathbb{R}^+ | w_0 \le \tau \le w_L\}$.

In fact, these results respectively provide w.h.p. an approximation for the LSF extremities because within three SDs most randomly generated samples will be accounted

$$
\begin{aligned}
f_0(\tau) &= f_W(\tau) \cdot \mathcal{N}_S(l, \sigma_\Psi^2) = \frac{2\sqrt{2}\ln(10) \cdot 10^{2(\tau-\alpha)/\beta}}{\sqrt{\pi}\left(3\sqrt{3}L^2 - 2\pi r_0^2\right)\beta\sigma_\Psi} \cdot \exp\left\{-(\tau-l)^2 / 2\sigma_\Psi^2\right\} \\
&\quad \times \left\{\pi \cdot \mathbf{1}(w_0 \le \tau \le w_I) - 2\pi \cdot \mathbf{1}(w_I \le \tau \le w_L) + 6\arcsin\left(\sqrt{3}L/2 \cdot 10^{(\tau-\alpha)/\beta}\right) \cdot \mathbf{1}(w_I \le \tau \le w_L)\right\} \\
&= \frac{2\sqrt{2}\ln(10) \cdot 10^{-2\alpha/\beta}}{\sqrt{\pi}\left(3\sqrt{3}L^2 - 2\pi r_0^2\right)\beta\sigma_\Psi} \cdot \exp\Big\{\overbrace{2\ln(10)\tau/\beta - (\tau-l)^2 / 2\sigma_\Psi^2}^{\triangleq q(\tau)}\Big\} \\
&\quad \times \left\{\pi \cdot \mathbf{1}(w_0 \le \tau \le w_I) - 2\pi \cdot \mathbf{1}(w_I \le \tau \le w_L) + 6\arcsin\left(\sqrt{3}L/2 \cdot 10^{(\tau-\alpha)/\beta}\right) \cdot \mathbf{1}(w_I \le \tau \le w_L)\right\}
\end{aligned} \tag{15}
$$

Next, we must integrate the expression of (15); however, this undertaking will require various intermediate steps which are respectively detailed in Appendix A.

As a reminder from (14), the $l$ entry represents a random sample of the LSF between a reference and an arbitrary terminal. Because of the log-normal nature of shadowing, this variable is expected to be in $\mathbb{R}$; yet from a practical standpoint, it is a.s. element in $\mathbb{R}^+$. For further precision, the range for this RV can additionally be narrowed-down. Indeed, the lower extremity of the LSF measure is analyzed in (16), where the optimization is split because the contributions from the average PL and shadowing are independent of each other. By the same token, the higher extremity for an $L$ size cellular network model is obtained in (17).

$$
\begin{aligned}
l_0 &\triangleq \min_{(r,\sigma_\Psi)\in\mathbb{R}_+^2}\left\{L_{PL}(r)_{dB}\right\} = \min_{r\in\mathbb{R}^+}\left\{\overline{L_{PL}(r)_{dB}}\right\} \\
&\quad + \min_{\sigma_\Psi\in\mathbb{R}^+}\left\{\Psi_{S-dB} \sim \mathcal{N}_S(0, \sigma_\Psi^2)\right\} \approx \tilde{l}_0 \\
&= \alpha + \beta\log_{10}(r_0) - 3\sigma_\Psi
\end{aligned} \tag{16}
$$

$$
l_L \triangleq \max_{(r,\sigma_\Psi)\in\mathbb{R}_+^2}\left\{L_{PL}(r)_{dB}\right\} \approx \tilde{l}_L = \alpha + \beta\log_{10}(L) + 3\sigma_\Psi \tag{17}
$$

for; i.e. with a confidence interval (CI) represented by: $\Pr\left\{\left|l - \overline{L_{PL}(r)_{dB}}\right| \le 3\sigma_\Psi\right\} \approx 0.997300$. Taken as a whole, we thus identify a tighter support range for $l$, given by:

$$
\left\{l \in \mathbb{R}^+ \;\middle|\; 0 < \tilde{l}_0 \lesssim l \lesssim \tilde{l}_L < \infty\right\} \tag{18}
$$

At this moment, we have all the necessary features to analytically assemble the PDF of the channel-loss. To be precise, from (A.3), we recognize that the density function is composed of two parts. The first part, which is designated by $K_0$, is identified in (A.2). The second part, namely $I_{LSF}(l)$, is obtained in (A.8), and its associated variables were solved in (A.9). Next, the domain of the density function was detailed in (18), where the related boundaries were assessed in (16) and (17). Finally, the exact closed-form stochastic statement for the PDF of the LSF between a randomly positioned node and a reference BS over a MCN model is explicitly shown in (19). Overall, the derived density result is generic due to the changeable parameters specified by the $\vec{\Lambda}$ array.

$$
\begin{aligned}
f_{L_{PL}}(l, \vec{\Lambda}) &= \left\{4 \cdot \ln(10) \cdot 10^{2(l-\alpha)/\beta} / \beta\left(3\sqrt{3}L^2 - 2\pi r_0^2\right)\right\} \cdot \exp\left\{\left(\sqrt{2} \cdot \ln(10)\sigma_\Psi/\beta\right)^2\right\} \\
&\quad \times \left\{\begin{aligned}
&\pi \cdot \left\{Q(z_0(l)) - 3 \cdot Q(z_I(l)) + 2 \cdot Q(z_L(l))\right\} \\
&+ 3\sqrt{2/\pi} \cdot \int_{z=z_I(l)}^{z_L(l)} \exp(-z^2/2) \cdot \arcsin\left(\sqrt{3}L \cdot 10^{-\sigma_\Psi z/\beta} / 2 \cdot 10^{\{\beta(l-\alpha)+2\ln(10)\sigma_\Psi^2\}/\beta^2}\right) dz
\end{aligned}\right\}
\end{aligned} \tag{19}
$$

- $\vec{\Lambda} = \{\alpha, \beta, \sigma_\Psi, r_0, L\} \in \mathbb{R}_+^5$ $\qquad 0 < \tilde{l}_0 \lesssim l \lesssim \tilde{l}_L < \infty$
- $Q(z) = erfc\left(z/\sqrt{2}\right)/2 = \left\{1 - erf\left(z/\sqrt{2}\right)\right\}/2$
- $z_0(l) = \left\{\alpha - l + \ln\left(r_0^{\beta/\ln(10)} / 10^{2\sigma_\Psi^2/\beta}\right)\right\} / \sigma_\Psi$
- $z_I(l) = \left\{\alpha - l + \ln\left(\left(\sqrt{3}L/2\right)^{\beta/\ln(10)} / 10^{2\sigma_\Psi^2/\beta}\right)\right\} / \sigma_\Psi$
- $z_L(l) = \left\{\alpha - l + \ln\left(L^{\beta/\ln(10)} / 10^{2\sigma_\Psi^2/\beta}\right)\right\} / \sigma_\Psi$
- $\tilde{l}_0 = \alpha + \beta\log_{10}(r_0) - 3\sigma_\Psi$ • $\tilde{l}_L = \alpha + \beta\log_{10}(L) + 3\sigma_\Psi$

# 5. EXPERIMENTAL VALIDATION BY MC SIMULATIONS

Here, we will authenticate the expression for the LSF distribution of (19) by means of stochastic simulations. Generally speaking, the approach for the validation process is broken-down into three major steps: (i) for a given lattice structure and dimensions, the random network geometry of wireless nodes is emulated via MC approach; (ii) the LSF density for a particular channel environment is numerically estimated using the emulated spatial samples; and (iii) the analytically derived PDF is plotted and then compared with the scholastic estimation.

It is imperative to emphasize that the tractable expression of (19) is fully generic and thus can be adaptable for any cellular application and wireless technology, as long as user's spatial geometry is assumed to be *random* and *uniform* over a MCN grid. Although the obtained result is generic in nature, yet to examine its correctness, we will exclusively consider the channel parameters of IEEE 802.20 [25] for an urban macrocell as specified in Table III. The actual details for the MC simulations are outlined as follows:

- In Table III, the transmission radius $L$ can take different values. We will however consider a cellular size of 600 m, which translates into an RCR of ~17.14. Given this RCR value, we therefore realize from (12) that Cartesian-based RNG is more efficient.
- An $n_S = 10,000$ random samples for nodes 2D spatial position is required. In fact, the set of $\hat{x}_i$ : $i = 1, 2, \cdots, n_S$ random components are generated from the algorithm of Figure 3. After, based on these values, the $\hat{y}_i$ counterparts are obtained using the approach described by (7).
- The distance $\hat{r}_i$ between the reference BS and random nodes is then calculated using the simple Pythagorean theorem.
- After that, the average PL for each of the $n_S$ random samples is computed by:

$$\hat{\bar{l}}_i \triangleq \overline{L_{PL}\left(\hat{r}_i\right)_{dB}} = \alpha + \beta \log_{10}\left(\hat{r}_i\right) \qquad i = 1, 2, \cdots, n_S \quad (20)$$

**Table III.** MBWA channel model for urban macrocell.

| IEEE 802.20 Propagation Parameters | |
|---|---|
| Propagation Model: | *COST*-231 *Hata-Model* |
| Operating Frequency: | 1.9 GHz |
| Support Range: | $r_0 = 35 \text{ m} \le r \le L$ <br> $600 \le L \le 3{,}500$ m |
| Channel-Loss: | $\alpha = 34.5$ dB <br> $\beta = 35$ dB |
| Shadowing: | $\sigma_\Psi = 10$ dB |

MBWA, mobile broadband wireless access.

- Next, values for shadowing are generated such that $\hat{n}_i \sim \mathcal{N}(0,1)$ are samples from a standard normal distribution in order to get instances of LSF as expressed by:

$$\hat{l}_i \triangleq L_{PL}\left(\hat{r}_i\right)_{dB} = \hat{\bar{l}}_i + \hat{\psi}_i = \hat{\bar{l}}_i + \sigma_\Psi \hat{n}_i \qquad i = 1, 2, \cdots, n_S \quad (21)$$

- The uppermost plot of Figure 10 shows a scatter diagram for the LSF as a function of the BS to node interpoint range. Specifically, each of the 10,000 instances is represented by a random point. For perspective to this MC realization, three deterministic plots, namely, $\overline{L_{PL}(r)_{dB}}$, $\overline{L_{PL}(r)_{dB}} - 3\sigma_\Psi$, and $\overline{L_{PL}(r)_{dB}} + 3\sigma_\Psi$ over $r \in [r_0, L]$ are also shown so as to characterize the average PL and the ~99.7% CI of LSF caused by shadowing. Indeed, as noticeable from the figure, only a negligible of ~0.3% of samples can be found outside the delineation of the CI.
- Then, based on the described scatter plot, a histogram for the LSF measure is constructed. In this simulation, an $n_B = 100$ bin histogram is considered with equal width designated by $\Delta l_B \in \mathbb{R}^+$. Precisely, the bars of the histogram are positioned next to each other with no spacing among them. As for the quantity of occurrence per bar, they are accordingly scaled to reflect an estimate of the PDF; that is, the occurrence is divided by the amount of random samples and the bin width. Once scaling is performed, we obtain the PDF estimation at discrete points, namely, $pdf_j : j = 1, 2, \cdots, n_B$.
- Also, the CDF of the LSF measure for randomly positioned nodes is approximated by the following recursive relationship:

$$\begin{aligned} cdf_1 &= pdf_1 \cdot \Delta l_B \\ cdf_j &= cdf_{j-1} + pdf_j \cdot \Delta l_B \qquad j = 2, 3, \cdots, n_B \end{aligned} \quad (22)$$

- As shown in Figure 10, the PDF estimation is performed over two values of $n_S$. As expected, an increase of random samples produces a better estimate that appropriately matches the theoretically derived density function of the LSF.
- We remarked earlier in Table III that the cellular size varies from $L = 600 \rightarrow 3{,}500$ m. Therefore, we find it intriguing to randomly simulate the LSF-PDF as $L$ changes. The result of this undertaking is shown in Figure 11. It is worth noting from the simulation that an increase in the cellular dimension raises the channel-loss interval, and as a result, the first-moment of the PDF is further shifted to the right. Also, it is obvious that the analytical derivation of the PDF and the estimation are properly congruent to each other.

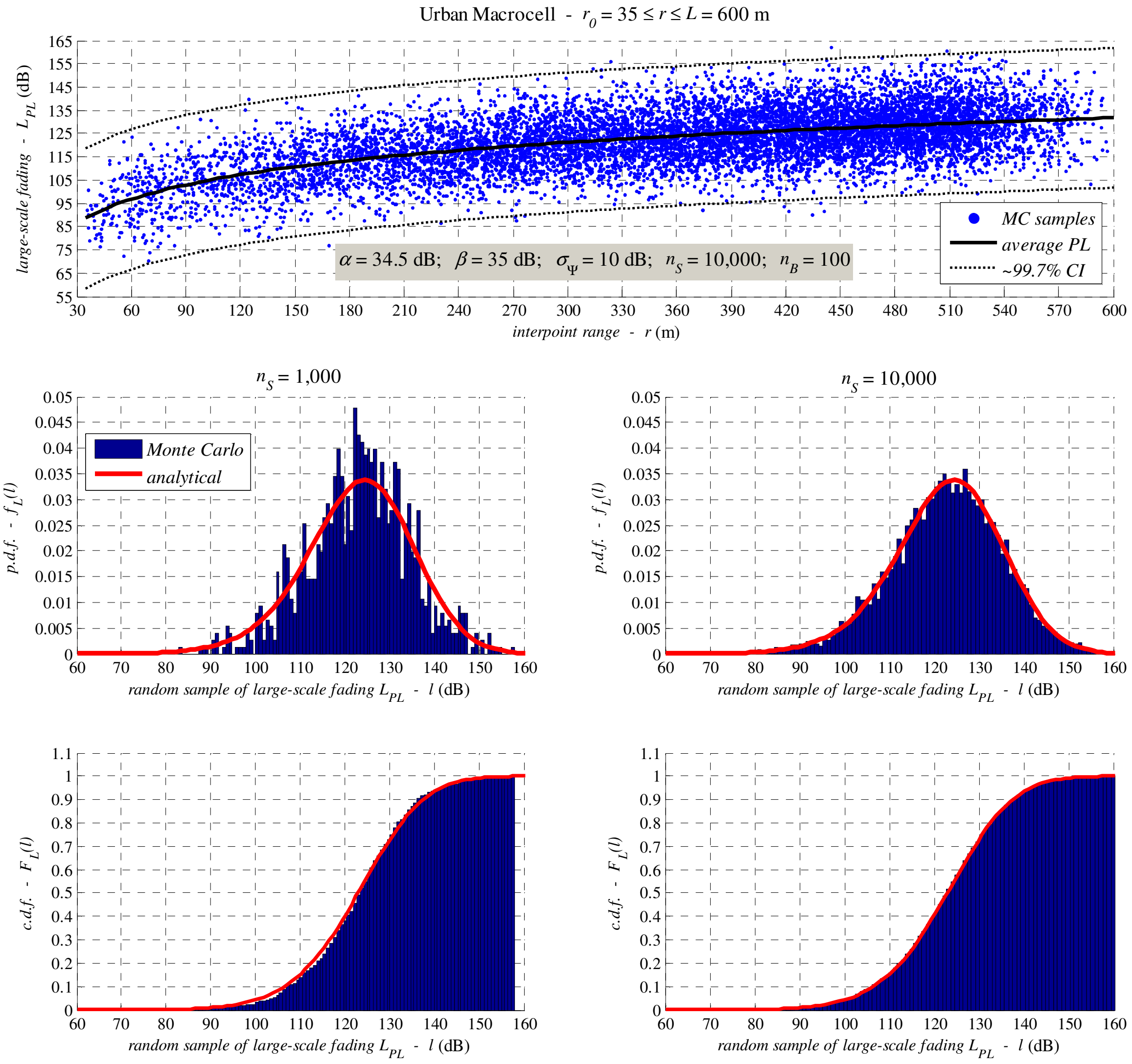

**Figure 10.** Verifying the analytically derived formulation for the large-scale fading distribution. PL, path-loss; CI, confidence interval; PDF, probability density function; CDF, cumulative distribution function.

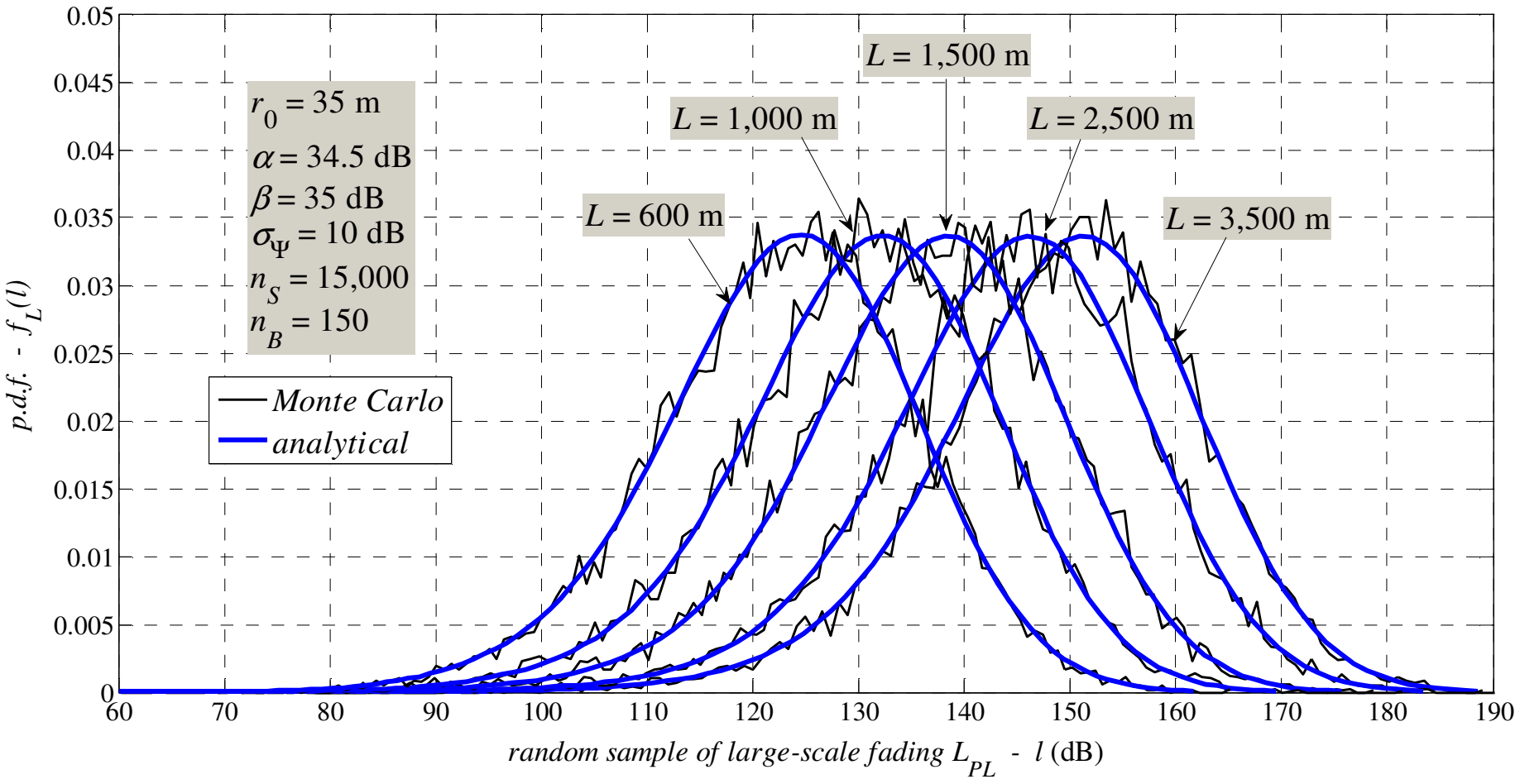

**Figure 11.** Large-scale fading distribution for centralized connectivity over different cellular sizes. PDF, probability density function.

# 6. CONCLUSION

The main objective of this paper was to describe the channel-loss density for a random network with respect to its service provider. In fact, such density can be obtained numerically using MC simulations. However this approach is computationally expensive, and also it does not produce a tractable and generic stochastic statement useful for analysis and interplay of input/output parameters. Consequently, in order to mathematically characterize with great precision the manifestation of the channel decay, we progressed into various technical steps.

In particular, we first had to explain the essential groundwork for the derivation of LSF density by specifying and combining the analytical features of the spatial homogeneity, the geometrical attributes of the MCN lattice, and the characteristics of radiation.

Next, we developed an efficient approach for emulating the geometry of the random MCN geared specifically for LSF analysis. This was performed as a preliminary step in deriving the LSF distribution and also for verifying the authenticity of the derivation via actual spatial deployment. We also measured the performance of the RNG, and its stochastic features were theoretically formulated and experimentally evaluated.

Equipped with all the necessary steps, we then analytically derived the exact and closed-form expression for the LSF density function between a prepositioned reference BS and a randomly deployed node. We then performed various MC simulations in order to ensure and confirm the veracity of the result. To be precise, in this derivation we took into account a number of fundamentally important elements, namely, the cellular structure of the architecture, the nodal spatial emplacement, the far-field effect of the reference antenna, the PL behavior, and the impact of channel scatterers.

In fact, the final and overall stochastic expression of the LSF-PDF expressed in (19), is entirely generic and can directly be adjusted to any cellular size $L$, close-in distance $r_0$, PL parameters $\alpha$ and $\beta$, and shadowing features described by its SD $\sigma_\Psi$. That is to say that the stochastic formulation was attained in such a way that it could be applied to numerous MCN applications and technologies having a particular scale, coverage, and channel features. In other words, as shown in Figure 12, the reported predictive result is adaptable via the insertion of related variables to the different network architectures, such as, femtocell, picocell, microcell, and macrocell systems.

Also, given the diversity of the transmission coverage for each of the listed network realizations, it is thus evident to recognize the variability of the RCR. Notably, for mobile applications that operate with microcell or macrocell networks, the RCR is generally in the order of ten or greater. As for femtocell and picocell communications, the RCR is typically smaller than this value. Therefore, when the RCR has a slighter level, the significance of the BS far-field radiation is more prominent. On the other hand, a superior RCR is marginally impacted by the far-field region. Nonetheless, this EM propagation phenomenon was explicitly considered in the derived density of the LSF model in order to characterize the laws of communications in a rigorous manner; and also to ensure the soundness of the stochastic expression for all type of cellular systems, irrespective of the network scheme.

Finally, as remarked in earlier parts of the paper, it is worthwhile to emphasize that the closed-form analytical expression of the channel-loss PDF will be applicative for all cellular network cases shown in Figure 13, irrespective of the considered sectoring type and the cluster rotation angle $\phi$.

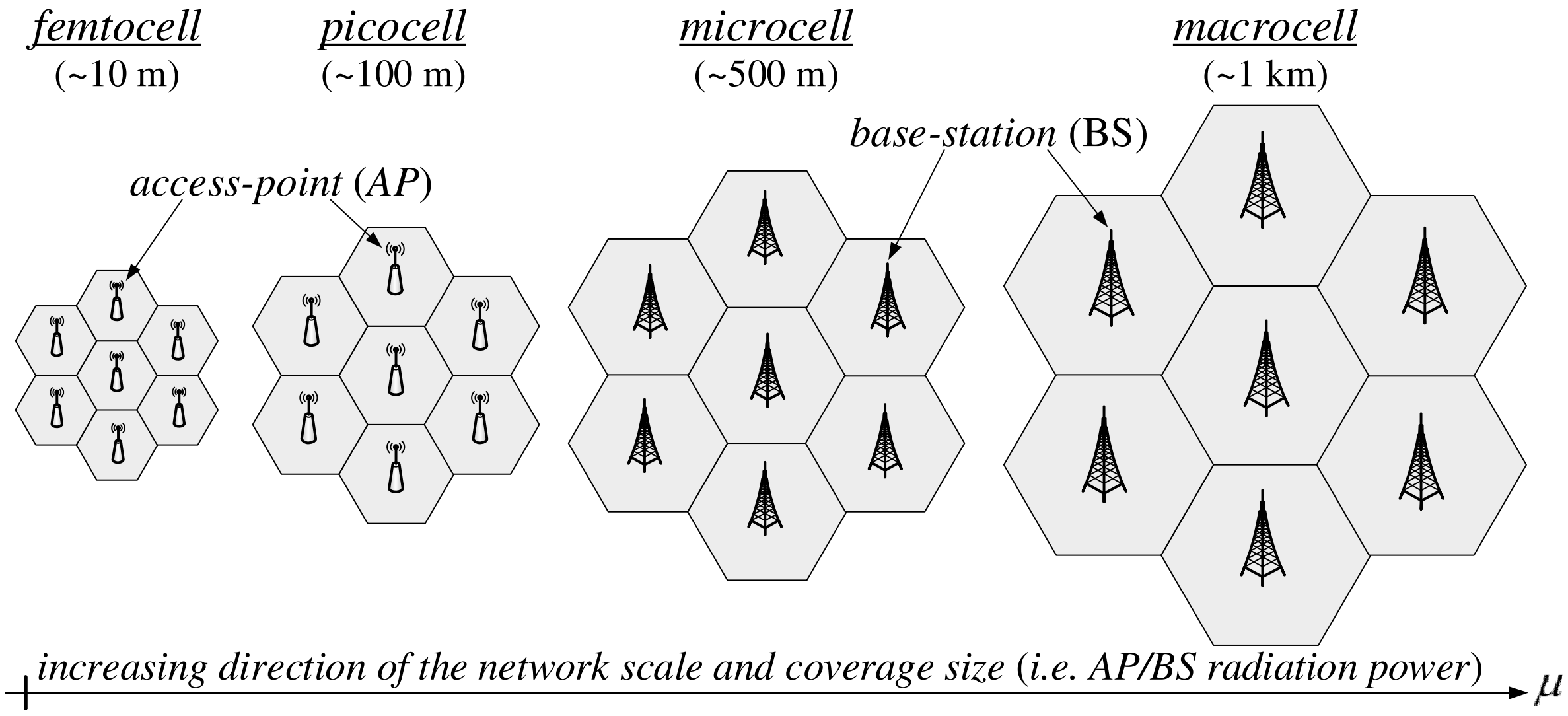

**Figure 12.** Feasibility of the multi-cellular network model for various deployment applications and purposes.

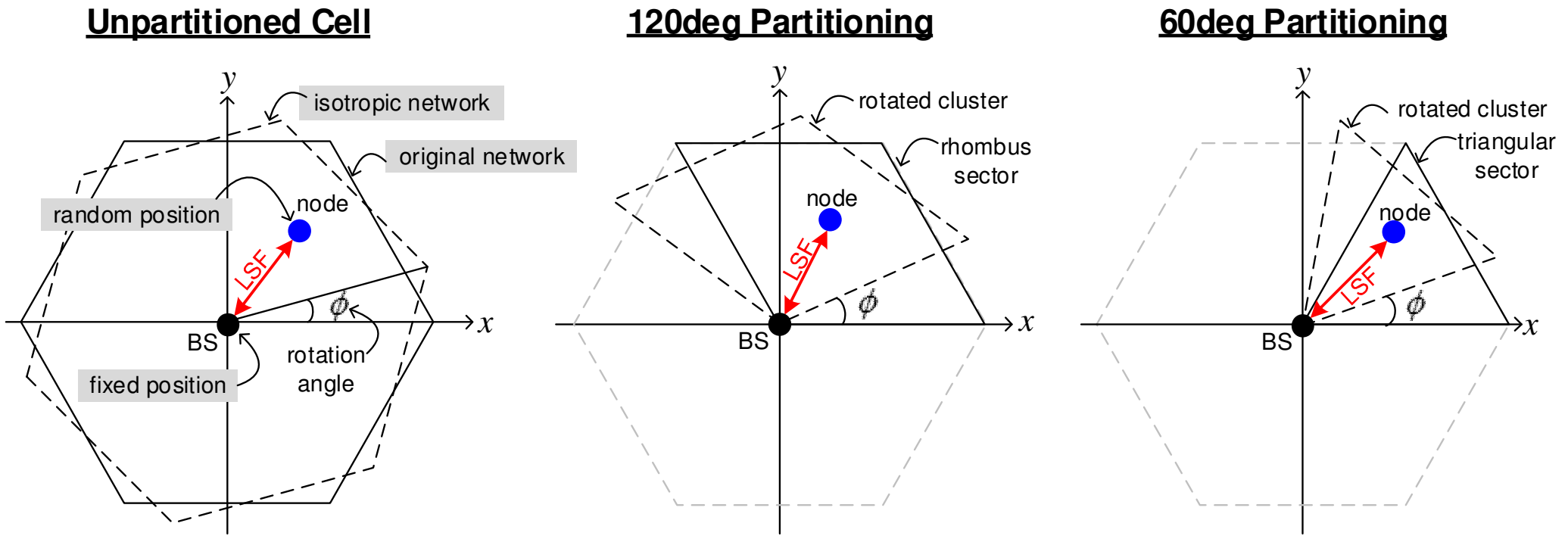

**Figure 13.** Applicability of the formulated large-scale fading (LSF) distribution for different random deployments. BS, base-station.

## APPENDIX A: INTEGRATING EQUATION (15)

First, we arrange (15) by completing the square of the quadratic function $q(\tau)$ inside the exponential so that it becomes of the form: $q(\tau) = a(\tau - h)^2 + k$. After some arithmetical manipulations, we then recognize that: $a = -1/2\sigma_\Psi^2$, $h = \{l + 2\ln(10)\sigma_\Psi^2/\beta\}$, and $k = \left\{2 \cdot l \cdot \ln(10)/\beta + \left(\sqrt{2}\ln(10)\sigma_\Psi/\beta\right)^2\right\}$. Now, the exponential part of (15) can be reorganized:

$$
\begin{aligned}
\exp\{q(\tau)\} &= \exp\left\{a(\tau-h)^2 + k\right\} = \exp\{k\}\cdot\exp\left\{a(\tau-h)^2\right\} \\
&= \exp\left\{2\cdot l\cdot\ln(10)/\beta + \left(\sqrt{2}\ln(10)\sigma_\Psi/\beta\right)^2\right\}\cdot\exp\left\{-\left(\tau-\{l+2\ln(10)\sigma_\Psi^2/\beta\}\right)^2/2\sigma_\Psi^2\right\} \\
&= 10^{2\cdot l/\beta}\cdot\exp\left\{\left(\sqrt{2}\ln(10)\sigma_\Psi/\beta\right)^2\right\}\cdot\exp\left\{-\left(\tau-\{l+2\ln(10)\sigma_\Psi^2/\beta\}\right)^2/2\sigma_\Psi^2\right\}
\end{aligned}
\tag{A.1}
$$

After substituting (A.1) into (15), we then find that:

$$
\begin{aligned}
f_0(\tau) = &\overbrace{\frac{2\sqrt{2}\ln(10)\cdot 10^{2(l-\alpha)/\beta}}{\sqrt{\pi}\left(3\sqrt{3}L^2 - 2\pi r_0^2\right)\beta\sigma_\Psi}\exp\left\{\left(\sqrt{2}\ln(10)\sigma_\Psi/\beta\right)^2\right\}}^{\triangleq K_0}\cdot\exp\left\{\frac{-1}{2\sigma_\Psi^2}\left(\tau-\{l+2\ln(10)\sigma_\Psi^2/\beta\}\right)^2\right\} \\
&\times\left\{\pi\cdot\mathbf{1}(w_0\le\tau\le w_l) - 2\pi\cdot\mathbf{1}(w_l\le\tau\le w_L) + 6\arcsin\left(\sqrt{3}L/2\cdot 10^{(\tau-\alpha)/\beta}\right)\cdot\mathbf{1}(w_l\le\tau\le w_L)\right\}
\end{aligned}
\tag{A.2}
$$

At present, the function in (A.2) is adequately ordered for the purpose of being integrated, where the $\tau$ independent expressions are assigned to $K_0$. Taken together, the LSF distribution of (14) can be split into three parts where each has a particular identifier:

$$
f_{L_{PL}}(l) = \int_{\tau=-\infty}^{\infty} f_0(\tau)\,d\tau = K_0\underbrace{\left\{\underbrace{\int_{\tau=w_0}^{w_l} f_0^{(1)}(\tau)\,d\tau}_{\triangleq I_{LSF}^{(1)}(l)} + \underbrace{\int_{\tau=w_l}^{w_L} f_0^{(2)}(\tau)\,d\tau}_{\triangleq I_{LSF}^{(2)}(l)} + \underbrace{\int_{\tau=w_l}^{w_L} f_0^{(3)}(\tau)\,d\tau}_{\triangleq I_{LSF}^{(3)}(l)}\right\}}_{\triangleq I_{LSF}(l)} \quad l\in\mathbb{R}
\tag{A.3}
$$

Following the first integration, we obtain (A.4), where $z = z(\tau) = \left(\tau - \{l + 2\ln(10)\sigma_\Psi^2/\beta\}\right)/\sigma_\Psi$; and $Q(z)$ is an alternate format of the complementary error function (ERFC).

$$
\begin{aligned}
I_{LSF}^{(1)}(l) &= \int_{\tau=w_0}^{w_l} f_0^{(1)}(\tau)\,d\tau \\
&= \pi\int_{\tau=w_0}^{w_l}\exp\left\{-\left(\tau-\{l+2\ln(10)\sigma_\Psi^2/\beta\}\right)^2/2\sigma_\Psi^2\right\}d\tau \\
&= \pi\cdot\sigma_\Psi\int_{z=z_0}^{z_l}\exp\left(-z^2/2\right)dz \\
&= -\pi\sqrt{2\pi}\cdot\sigma_\Psi\cdot\{Q(z)\}_{z=z_0}^{z_l}
\end{aligned}
\tag{A.4}
$$

The second integration of (A.3) is relatively similar to (A.4), and so it can readily be solved as follows:

$$
I_{LSF}^{(2)}(l) = 2\pi\sqrt{2\pi}\cdot\sigma_\Psi\cdot\{Q(z)\}_{z=z_l}^{z_L}
\tag{A.5}
$$

At this point, we could get an intermediate result by adding (A.4) and (A.5) together:

$$I_{LSF}^{(1)}(l)+I_{LSF}^{(2)}(l)=\pi\sqrt{2\pi}\sigma_\Psi\left\{2\{Q(z)\}_{z=z_I}^{z_L}-\{Q(z)\}_{z=z_0}^{z_I}\right\}$$
$$=\pi\sqrt{2\pi}\sigma_\Psi\{Q(z_0)-3Q(z_I)+2Q(z_L)\} \tag{A.6}$$

As for the third integration defined in (A.3), it is manifested by:

$$I_{LSF}^{(3)}(l)=\int_{\tau=w_I}^{w_L} f_0^{(3)}(\tau)\,d\tau=6\int_{\tau=w_I}^{w_L}\exp\left\{-\left(\tau-\{l+2\ln(10)\sigma_\Psi^2/\beta\}\right)^2\Big/2\sigma_\Psi^2\right\}\cdot\arcsin\left(\frac{\sqrt{3}L}{2\cdot 10^{(\tau-\alpha)/\beta}}\right)d\tau$$
$$=6\cdot\sigma_\Psi\int_{z=z_I}^{z_L}\exp\left(-z^2/2\right)\cdot\arcsin\left(\sqrt{3}L\cdot 10^{-\sigma_\Psi z/\beta}\Big/2\cdot 10^{\{\beta(l-\alpha)+2\ln(10)\sigma_\Psi^2\}/\beta^2}\right)dz \tag{A.7}$$

If we combine the results of (A.6) and (A.7) together, we then get the notation in (A.8), where and $z=z(w(r))$ is a composed function of PL and geometrical separation as detailed by (A.9).

$$I_{LSF}(l)=\sigma_\Psi\left\{\begin{aligned}&\pi\sqrt{2\pi}\cdot\{Q(z_0)-3Q(z_I)+2Q(z_L)\}\\&+6\cdot\int_{z=z_I}^{z_L}\exp\left(-z^2/2\right)\cdot\arcsin\left(\sqrt{3}L\cdot 10^{-\sigma_\Psi z/\beta}\Big/2\cdot 10^{\{\beta(l-\alpha)+2\ln(10)\sigma_\Psi^2\}/\beta^2}\right)dz\end{aligned}\right\} \tag{A.8}$$

$$\begin{aligned}z_{0,I,L}=z_{0,I,L}(l)&=\left(\tilde{w}-\{l+2\ln(10)\sigma_\Psi^2/\beta\}\right)\Big/\sigma_\Psi \quad\longleftrightarrow\quad \tilde{w}=w_0,w_I,w_L\\&=\left(\alpha+\beta\log_{10}(\tilde{r})-\{l+2\ln(10)\sigma_\Psi^2/\beta\}\right)\Big/\sigma_\Psi \quad\longleftrightarrow\quad \tilde{r}=r_0,\sqrt{3}L/2,L\\&=\left\{\alpha-l+\ln\left(\tilde{r}^{\beta/\ln(10)}\Big/10^{2\sigma_\Psi^2/\beta}\right)\right\}\Big/\sigma_\Psi\end{aligned} \tag{A.9}$$

## REFERENCES

1. Rappaport TS. *Wireless Communications: Principles and Practice*. Prentice Hall PTR: New Jersey, 2002.
2. Goldsmith A. *Wireless Communications*. Cambridge University Press: New York, 2005.
3. Saunders SR. *Antennas and Propagation for Wireless Communication Systems*. John Wiley & Sons, 1999.
4. Seybold JS. *Introduction to RF Propagation*. Wiley: New Jersey, 2005.
5. Garg V. *Wireless Communications and Networking*. Morgan Kaufmann Publishers: California, 2007.
6. Bharucha Z, Haas H. The distribution of path losses for uniformly distributed nodes in a circle. *Research Letters in Communications* 2008; **2008**: 1–4.
7. Broyde Y, Messer H. A cellular sector-to-users path loss distribution model. In *Proc. of the 15th IEEE/SP Workshop on Statistical Signal Processing (SSP'09)*, Cardiff, UK, 2009; 321–324.
8. Abdulla M, Shayan YR, Baek JH. Revisiting circular-based random node simulation. In *Proc. of the 9th IEEE International Symposium on Communication and Information Technology* (*ISCIT'09*), Incheon, South Korea, Sep. 2009; 731–732.
9. Abdulla M, Shayan YR. An exact path-loss density model for mobiles in a cellular system. In *Proc. of the 7th ACM International Symposium on Mobility Management and Wireless Access (*MobiWac'09*), held in conjunction with MSWiM'09*, Tenerife, Canary Islands, Spain, Oct. 2009; 118–122.
10. Baltzis KB. Analytical and closed-form expressions for the distribution of path loss in hexagonal cellular networks. *Wireless Personal Communications* 2011; **60**(4): 599–610.
11. Abdulla M, Shayan YR. Closed-form path-loss predictor for Gaussianly distributed nodes. In *Proc. of IEEE International Conference on Communications* (*ICC'10*), Cape Town, South Africa, May 2010; 1–6.
12. Abdulla M. Simple subroutine for inhomogeneous deployment. In *Proc. of the 6th IEEE Global Information Infrastructure and Networking Symposium* (*GIIS'14*), Montréal, Québec, Canada, Sep. 2014; 1–3.
13. Jakó Z, Jeney G. Outage probability in Poisson-cluster-based LTE two-tier femtocell networks. *Wiley's Wireless Communications and Mobile Computing Journal* 2014: 1–12, doi: 10.1002/wcm.2485.
14. Dong L, Petropulu AP, Poor HV. A cross-layer approach to collaborative beamforming for wireless ad hoc networks. *IEEE Trans. on Signal Processing* 2008; **56**(7): 2981–2993.
15. Omiyi P, Haas H, Auer G. Analysis of TDD cellular interference mitigation using busy-bursts. *IEEE Trans. on Wireless Communications* 2007; **6**(7): 2721–2731.

16. Mukherjee S, Avidor D, Hartman K. Connectivity, power, and energy in a multihop cellular-packet system. *IEEE Trans. on Vehicular Technology* 2007; **56**(2): 818–836.
17. Ochiai H, Mitran P, Poor HV, Tarokh V. Collaborative beamforming for distributed wireless ad hoc sensor networks. *IEEE Trans. on Signal Processing* 2005; **53**(11): 4110–4124.
18. Bettstetter C. On the connectivity of ad hoc networks. *Computer Journal* 2004; **47**(4): 432–447.
19. Abdulla M, Shayan YR. Cellular-based statistical model for mobile dispersion. In *Proc. of the 14th IEEE International Workshop on Computer-Aided Modeling, Analysis and Design of Communication Links and Networks* (*CAMAD'09*), Pisa, Tuscany, Italy, Jun. 2009; 1–5.
20. Abdulla M. On the Fundamentals of Stochastic Spatial Modeling and Analysis of Wireless Networks and its Impact to Channel Losses. *Ph.D. Dissertation*, Dept. of Electrical and Computer Engineering, Concordia Univ., Montréal, Québec, Canada, September 2012.
21. Krishnan K. *Probability and Random Processes.* Wiley-Interscience: New Jersey, 2006.
22. Mathai AM. *An Introduction to Geometrical Probability: Distributional Aspects with Applications.* CRC Press: Amsterdam, The Netherlands, 1999.
23. Srinivasa S, Haenggi M. Distance distributions in finite uniformly random networks: theory and applications. *IEEE Trans. on Vehicular Technology* 2010; **59**(2): 940–949.
24. Khalid Z, Durrani S. Distance distributions in regular polygons. *IEEE Trans. on Vehicular Technology* 2013; **62**(5): 2363–2368.
25. IEEE 802.20 channel models document. *IEEE 802.20 Working Group on Mobile Broadband Wireless Access* 2007: 1–33.

## AUTHORS' BIOGRAPHIES

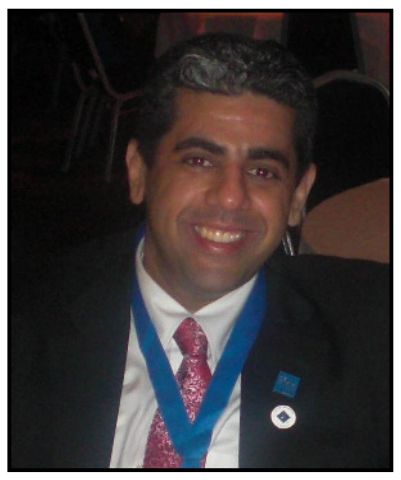

**Mouhamed Abdulla** received, respectively in 2003, 2006, and 2012, a BEng degree (with Distinction) in Electrical Engineering, an MEng degree in Aerospace Engineering, and a PhD degree in Electrical Engineering all at Concordia University in Montréal, Québec, Canada. He is currently an NSERC Postdoctoral Research Fellow with the Department of Electrical Engineering of the University of Québec. Previously, he was a Systems Engineering Researcher in the Wireless Design Laboratory of the Department of Electrical and Computer Engineering of Concordia University. Moreover, for nearly 7 years since 2003, he worked at IBM Canada Ltd. as a Senior Technical Specialist. Dr. Abdulla holds several awards, honors and recognitions from international organizations, academia, government, and industry. He is professionally affiliated with IEEE, IEEE ComSoc, IEEE YP, ACM, AIAA, and OIQ. Currently, he is a member of the IEEE Executive Committee of the Montréal Section, where he was the Secretary in 2013, and is presently the Treasurer of the Section. In addition, he is the Secretary of IEEE ComSoc and ITSoc societies. Furthermore, he is an Associate Editor of *IEEE Technology News Publication*, *IEEE AURUM Newsletter*; and Editor of *Elsevier Digital Communications and Networks*, *Journal of Next Generation Information Technology*, and *Advances in Network and Communications Journal*. He regularly serves as a referee for a number of Canadian Granting Agencies and Journal Publications such as: IEEE, Wiley, IET, EURASIP, Hindawi, Elsevier, and Springer. He also contributes as an examination writer for the *IEEE/IEEE ComSoc WCET® Certification Program*. Besides, he constantly serves as a Technical Program Committee (TPC) member for several international IEEE and Springer conferences. His research interest include: mobile communications, space/satellite communications, network performance, channel characterization and interference mitigation. Moreover, he has a particular interest in philosophical factors related to engineering education and research innovation. Since 2011, his biography is listed in the distinguished Marquis Who's Who in the World publication.

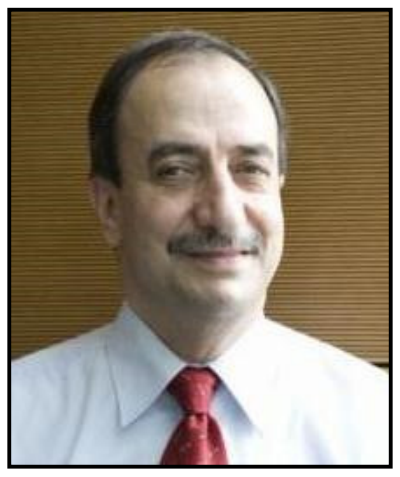

**Yousef R. Shayan** received his PhD degree in electrical engineering from Concordia University in 1990. Since 1988, he has worked in several wireless communication companies in different capacities. He has worked in research and development departments of SR Telecom, Spar Aerospace, Harris and BroadTel Communications, a company he co-founded. In 2001, Dr. Shayan joined the Department of Electrical and Computer engineering of Concordia University as associate professor. Since then he has been Graduate Program Director, Associate Chair and Department Chair. Dr. Shayan is founder of "*Wireless Design Laboratory*" at the Dept. of ECE which was established in 2006 based on a major CFI Grant. This lab has state-of-the art equipment which is used for development of wireless systems. In June 2008, Dr. Shayan was promoted to the rank of professor and was also recipient of "*Teaching Excellence Award*" for academic year 2007-2008 awarded by Faculty of Engineering and Computer Science. His fields of interest include: wireless communications, error control coding and modulation techniques.